%% file: main.tex
\documentclass[twocolumn]{IEEEtran}
\pdfoutput=1
\usepackage{graphics, theorem, subfigure, times, amsfonts, amsmath, amssymb, cite}
\usepackage{pgfplots}
\usepackage{color}
\usepackage{url}
\definecolor{darkblue}{rgb}{0,0.08,0.4} 
\definecolor{darkred}{rgb}{0.4,0.08,0} 
\input{mysymbol.sty}
\usepackage[margin=0.75in]{geometry}
\usepackage{multirow}
\usepackage{booktabs}

\newtheorem{proposition}{Proposition}
\newtheorem{corollary}{Corollary}
{}
\newtheorem{lemma}{Lemma}
\newtheorem{definition}{Definition}

{\itshape}{\rmfamily}

\makeatletter
\def\blfootnote{\xdef\@thefnmark{}\@footnotetext}
\makeatother

\usepackage{latexsym}
\usepackage{amscd, verbatim}
\usepackage{hyperref,url,setspace}
\usepackage{algorithm,algorithmic}

\title{Demand Response with Communicating Rational Consumers}

\author{Ceyhun Eksin, Hakan Deli\c{c} and Alejandro Ribeiro}

\date{\today}
\begin{document}
\normalsize
\maketitle

\begin{abstract}
The performance of an energy system under a real-time pricing mechanism depends on the consumption behavior of its customers, which involves uncertainties. In this paper, we consider a system operator that charges its customers with a real-time price that depends on the total realized consumption. Customers have unknown and heterogeneous consumption preferences. We propose behavior models in which customers act selfishly, altruistically or as welfare-maximizers. In addition, we consider information models where customers keep their consumption levels private, communicate with a neighboring set of customers, or receive broadcasted demand from the operator. Our analysis focuses on the dispersion of the system performance under different consumption models. To this end, for each pair of behavior and information model we define and characterize optimal rational behavior, and provide a local algorithm that can be implemented by the consumption scheduler devices. Analytical comparisons of the two extreme information models, namely, private and complete information models, show that communication model reduces demand uncertainty while having negligible effect on aggregate consumer utility and welfare. In addition, we show the impact of real-time price policy parameters have on the expected welfare loss due to selfish behavior affording critical policy insights. 
\end{abstract}

\section{Introduction}\blfootnote{C. Eksin is with the School of Electrical and Computer Engineering and the School of Biological Sciences at the Georgia Institute of Technology, Atlanta, GA. H. Deli\c{c} is with the Wireless Communications Laboratory, Department of Electrical and Electronics Engineering, Bo\u{g}azi\c{c}i University, Bebek 34342 Istanbul, Turkey. A. Ribeiro is with the Department of Electrical and Systems Engineering, University of Pennsylvania, Philadelphia, PA. Work supported by NSF CAREER CCF-0952867, NSF CCF-1017454, and the Bo\u{g}azi\c{c}i University Research Fund under Grant 13A02P4. }

Demand response management (DRM) emerges as a prominent method to alleviate the complications in power balancing caused by uncertainties both on the consumer and the supply side. Changes in user consumption preferences create the uncertainty on the consumer side while the uncertainty on the supply side is due to renewable resources. DRM refers to the system operator's effort to improve system performance by shaping consumption through pricing policies. Smart meters that can control the power consumption of customers, and enable information exchange between meters and the system operator (SO) provide the infrastructure to implement these policies. 

Real-time pricing (RTP) is a pricing policy where the price depends on instantaneous consumption of the population \cite{Autonomous_DSM_SG,SmartGrid_mechanism_design,Jarmo_et_al}. In RTP, the SO shares part of the risk and reward with its customers by setting price based on the total consumption. In these models, it is natural to propose game-theoretic models of consumption behavior, where users strategically reason about the behavior of others to anticipate price and determine their individual consumption \cite{Autonomous_DSM_SG, SmartGrid_mechanism_design,Jarmo_et_al,Xu_Schaar_2015,Li_et_al_2011, Atzeni_et_al, Yang_et_al, Depuru_et_al_2011}. The specifics of the behavior model and the information available impact the system welfare and is critical in assessing the benefits or disadvantages of a pricing scheme \cite{Depuru_et_al_2011,Wang_et_al_2014}. Given an RTP mechanism, our goal in this paper is to characterize rational price-anticipatory behavior models under different information exchange schemes, and comparatively assess their impact on system performance measures.

We consider an RTP scheme in which customers agree to a price function that increases linearly with total consumption and that depends on an unknown renewable energy parameter  (Section \ref{energy_provider_model}). The individual customer utility at each time depends on the individual's consumption preference and price both of which are in general unknown to others (Section \ref{users_payoffs_behavior}). Initially, the SO sends public information on its estimate of population's consumption preferences and renewable source generation. Customers use the public information and their self-preferences to anticipate total consumption and renewable source's effect on price, and respond rationally by consuming according to a Bayesian Nash equilibrium (BNE) strategy. In \cite{Eksin_et_al_15}, based on this energy market model, we propose and show the effectiveness of a peak-to-average ratio (PAR)-minimizing pricing strategy. 
 
In this paper we explore the effects of different consumer behavior models, where consumers respond rationally regarding their individual utility, the population's aggregate utility or the welfare (Section \ref{behavior_model_section}). As time progresses, past consumption decisions contain information about the preferences of others which individuals can use to make more informed decisions in the current time. Based on this observation, we propose three information exchange models, namely, private, action-sharing and broadcast (Section \ref{information_exchange_section}). In the private model, users do not receive any information besides the initial public signal by the SO. In action-sharing there exists a communication network on which users exchange their latest consumption decisions with their immediate neighbors. In broadcasting, the SO broadcasts the total consumption after each time step. We assume that the customer's power control scheduler can adjust the load consumption between time steps according to its preferences and information. That is, we are interested in modeling consumption behavior for shiftable appliances, e.g., electric vehicles, electronic devices, air conditioners, etc.  \cite{Roozbehani}. We formulate each consumer behavior model and information exchange model pair as a repeated game of incomplete information and characterize BNE behavior (Section \ref{grid_communication_section}). We use the explicit characterization to rigorously analyze the effects of each pair of behavior and information exchange model on demand, aggregate user utility and welfare respectively in Sections \ref{sec_effects_demand}-\ref{sec_effects_welfare}. These sections present analytical derivations for the results shown numerically in \cite{EksinEtal15c_a}. In addition, we provide additional simulation results on sensitivity of the performance metrics with respect to the renewable energy parameter term in the price.

Our findings can be summarized as follows. While providing more information to the consumers through action-sharing or broadcasting models does not have a significant impact on the expected aggregate utility or on the welfare, which is defined as the sum of aggregate utility and the operator's net revenue, it reduces the uncertainty in total demand. Action-sharing and broadcasting information exchange models eventually achieve the expected utility under complete information when the communication network is connected. Furthermore, while in a private information model increased correlation among preferences increases the demand uncertainty, in the broadcast model the demand variance tends to decrease with increasing correlation. When we consider the effects of behavior models --selfish, altruistic, or welfare maximizers,-- we show that altruistic behavior achieves the highest expected aggregate utility and welfare maximizers achieve the highest expected welfare. Furthermore, we characterize the expected improvement in the considered performance metric with respect to selfish behavior model in terms of pricing parameters. Finally, we show that increased correlation among user preferences tends to adversely affect aggregate utility and welfare. We discuss the possible policy implications of these findings for the operator and policy makers in Section \ref{discussion_section}.

\section{Demand Response Model} \label{problem_formulation}
%
\input{problem_formulation.tex}

\section{Bayesian Nash equilibria} \label{equilibrium_section}
%
\input{stage_game_solution.tex}
%
\section{Consumers' Bayesian Game}\label{grid_communication_section}
%
\input{communication.tex}

%
\input{simulation.tex}

\section{Discussion} \label{discussion_section}

We proposed a demand response management model based on the real-time pricing scheme where an operator responsible for supplying electricity to a set of consumers employed real-time demand dependent price. The pricing function is such that it linearly increases with total consumption per capita and decreases with increasing renewable energy generation in the time slot. This and other RTP policies proposed in the literature \cite{Huang_et_al_2015} allow the operator to shape consumer behavior in order to improve system level performances, e.g., demand uncertainty reduction, welfare maximization, PAR minimization, etc. However, the implications and the extent of the success of the RTP policies depend on consumer behavior, as well as information consumers have, which are unknown to the SO. This paper provided an extensive analysis of the sensitivity of the system performance measures to consumer behavior and information under an RTP mechanism. Our results illustrated that communication among consumers reduced demand uncertainty and showed the extent to which the flexibility that an RTP mechanism provides in managing shiftable demand of the consumers could impact system performance. 

From the perspective of the SO, giving information to the consumers is beneficial in that it reduces demand uncertainty which can reduce the conventional energy reserves held for worst case scenarios. Furthermore, our results on sensitivity of demand and welfare on the renewable energy term in the price provide clear tradeoffs to be considered by the SO. On one hand, renewable energy term can shape shiftable consumption according to the abundance or the scarcity of renewables. On the other hand, these manipulations cause significant changes to the aggregate user utility. From the perspective of a regulator, who is responsible for the well-being of the system and consumers, the relationship established between the expected aggregate consumer utility when users are altruistic and the expected aggregate consumer utility when users are selfish provides insight on the harm caused by the RTP mechanism parameters to the consumers. Moreover, the relationship between the expected welfares when users are selfish versus welfare-maximizing provides a guideline on how to select pricing policy parameters based on system cost parameters to maximize expected welfare. These insights can guide limitations that the regulator can put to the SO's RTP policy. Lastly, the illustrated positive effects of giving additional information to the consumers is a push toward investing in a communication network of smart meters among consumers. From the perspective of consumers that are self-interested, these results show that sharing their information will not have adverse effects in expected utility.

It should be observed that the aforementioned implications depend on specific modeling choices, namely, on the demand side the assumption of Bayesian information processing, and on the supply side the use of a quadratic form for the SO's cost. These choices may be simplistic. But in the following we claim that the results outlined here still provide meaningful guidelines if these restrictions are lifted. On the demand side, with regard to the model of Bayesian information processing, we remark that this is a benchmark model of information processing, and in our analysis we consider the two extremes of consumer information access, namely, private and complete information. The analysis of the two extreme cases provide the range of results we can expect from the performance measures in an information sharing consumer model. Hence, if we are to consider another information sharing model that is not Bayesian and has a lower computational complexity, e.g. \cite{Forouzandehmehr_et_al_2013}, we would expect our insights to be similar to the action-sharing model considered here as long as the proposed model aggregates information approximately correctly. As for the use of quadratic energy costs on the supply side, it is better to consider a model in which the cost for each device can be modeled as a linear function of the power dispatched from each device. In this case the cost model is an increasing piecewise linear function of total consumption as power is dispatched from more costly generators with increasing total consumption \cite{Papavasiliou_Oren_2014}. The quadratic cost function is a tractable approximation for the piecewise linear cost function and captures the fundamental property that higher energy production requires dispatching from more costly sources. The quantitative specifics may change for piecewise linear functions but the qualitative conclusions will be similar.

A prominent feature of the vision for the electricity grid of the future is that consumers play an active role in balancing supply and demand through  communication with the operator and amongst the consumers themselves. Furthermore, the grid of the future should be cleaner by allowing for increased renewable energy penetration. In order to attain this vision, we need to address two main challenges. First, making consumers active entities in balancing demand and supply can have unintended consequences such as increasing uncertainty in the system, and thus inflating environmental and monetary costs \cite{Depuru_et_al_2011,Wang_et_al_2014}. Second, we have to address the fact that renewables are inherently intermittent and hard to predict, and hence, if not dealt with systematically, can increase the need for conventional energy reserves defying their primary purpose. This paper took a step in the direction to overcome these challenges in order to realize the vision for the electricity grid. 

\input{main.bbl}

\end{document}

%% file: problem_formulation.tex

There are $N$ customers, each equipped with a power consumption scheduler. Individual power consumption of $i \in \ccalN:=\{1,\dots, N\}$ at time $h \in \ccalH := \{1,\dots, H\}$ is denoted by $l_{i,h}$. The total power consumed by  $N$ customers at time $h$ is $L_h := \sum_{i \in \ccalN} l_{i,h}$. 

\subsection{Real-time pricing}\label{energy_provider_model}

The SO implements an adaptive pricing strategy whereby customers are charged a slot-dependent price $p_h$ that varies linearly with the total power consumption $L_h$. The SO has a set of renewable source plants at its dispatch and incorporates renewable generation into the pricing strategy by a random renewable power term $\omega_h \in \reals$ that depends on the amount of renewable power produced at time slot $h$. The per-unit power price in time slot $h$ is set as 
\begin{align} \label{price}
   p_{h}(L_h; \omega_h) = \frac{\gamma_h}{N} (L_h + \omega_h)
\end{align}
where $\gamma_h >0$ is a policy parameter to be determined by the SO based on its objectives. The random variable $\omega_h$ is such that $\omega_h=0$ when renewable sources operate at their nominal benchmark capacity $\bar{W}_h$. If the realized production exceeds this benchmark, $W_h > \bar{W}_h$, the SO agrees to set $-L_h<\omega_h<0$ to discount the energy price and to share its revenue from the windfall. If the realized production is below benchmark, i.e., $W_h < \bar{W}_h$, the SO sets $\omega_h>0$ to reflect the additional charge on the customers. The specific dependence of $\omega_h$ on the realized energy production and the policy parameter, $\gamma_h$, are part of the supply contract between the SO and its customers. 

We assume that the SO uses a model on the renewable power generation -- see, e.g.,\cite{Mohsenian_et_al_2011} for the prediction of wind generation -- to estimate the value of $\omega_h$ at the beginning of time slot $h$. The mean estimate $\bar\omega_h :=E_{\omega_h}[\omega_h]$ of the corresponding probability density function $P_{\omega_h}$ is made available to all customers prior to the time slot. We do not make any assumptions on the accuracy or the structure of these forecasts. Inclusion of a renewable dependent term in the price functions allows the operator to use the flexibility of consumption behavior to compensate for peaks in conventional energy reserves caused by intermittent renewable generation \cite{Gan_et_al,Sioshansi_Short_2009,Papavasiliou_Oren_2014,Mohsenian_et_al_2011}. We discuss its possible implications further in Section \ref{discussion_section}.

The operator's price function maps the amount of energy demanded to the market price.  Observe that the price $p_{h}(L_h; \omega_h)$ at time $h$ becomes known {\it after} the end of the time slot. This is because price value depends on the total demand $L_h$ and the value of $\omega_h$ which are unknown a priori. The SO can employ the pricing policy in \eqref{price} to achieve certain system performances, e.g., minimizing PAR, maximizing welfare, etc., by picking its policy parameter $\gamma_h >0$ \cite{Eksin_et_al_15}.

\subsection{Power consumer} \label{users_payoffs_behavior}

User $i$'s consumption at time slot $h$, $l_{i,h}$, depends on his consumption preference $g_{i,h}>0$, modeled as a random variable that may vary across time slots. When user $i$ consumes $l_{i,h}$, its consumption utility increases linearly with its preference $g_{i,h}$ and decreases quadratically with a constant term $\alpha_h$, described as  $g_{i,h} l_{i,h} - \alpha_h l_{i,h}^2$. The utility of $i$ at time slot $h \in \ccalH$ is then captured by the difference between the consumption utility of $i$ and the monetary cost of consumption $l_{i,h}p_{h}(L_h; \omega_h)$:
\begin{align}\label{utility_i}
   u_{i,h}(l_{i,h}, L_{h} ; g_{i,h},\omega_h)
        =  - l_{i,h} p_{h}(L_h; \omega_h) 
            + g_{i,h} l_{i,h} 
            - \alpha_{h}  l_{i,h}^2.
\end{align}
Note that even if the SO's policy parameter is set to $\gamma_h=0$, the utility of user $i$ is maximized by $l_{i,h} = g_{i,h} / 2 \alpha_h$. 
Note that we choose $\alpha_h$ to be homogeneous among the consumers. Our results extend to the case where the constant $\alpha_h$ is heterogeneous. 

The utility of user $i$ depends on the total power, $L_h$, consumed at $h$, which implies that it depends on the powers  that are consumed by other users in the current slot, denoted by $l_{-i,h} := \{l_{j,h}\! : j\in \ccalN \setminus i\}$. Power consumption of others, $l_{-i,h}$, depends partly on their respective self-preferences, i.e., preferences $g_{-i,h}:=\{g_{j,h}\}_{j \neq i}$, which are, in general, {\it unknown} to user $i$. We assume, however, that there is a probability density function $P_{\bbg_h}(\bbg_h)$ on the vector of self-preferences $\bbg_h := [g_{1,h} ~ \ldots~ g_{N,h}]^T$ from which these preferences are drawn. We further assume that $P_{\bbg_h}$ is normal with mean $\bar{g}_h \bbone$ where $\bar{g}_h >0$ and $\bbone$ is an $N\times 1$ vector with one in every element, and covariance matrix $\bbSigma_h$:
\begin{align} \label{preference_distribution}
    P_{\bbg_h} (\bbg_h) 
          =   \ccalN\Big(\bbg_h;\ \bar{g}_h \bbone, \bbSigma_h\Big) .
\end{align}
We use the operator $E_{\bbg_h}[\cdot]$ to signify expectation with respect to $P_{\bbg_h}$ and $\sigma^h_{ij}$ to denote the $(i,j)$th entry of the covariance matrix $\bbSigma_h$. Having mean $\bar{g}_h \bbone$ implies that all customers have equal average preferences in that $E_{\bbg_h}[g_{i,h}] = \bar{g}_h$ for all $i$. If $\sigma_{ij}^h=0$ for some pair $i\neq j$, it means that the self-preferences of these customers are uncorrelated. In general, $\sigma_{ij}^h\neq 0$ to account for correlated preferences due to, e.g., common weather. We assume that if there is a change in the consumption preferences from one time slot to the other, then the self-preferences $\bbg_h$ and $\bbg_{k}$ for different time slots $h\neq k$ are independent.

At the beginning of time slot $h$, we assume that $P_{\bbg_h}$ in \eqref{preference_distribution} is correctly predicted by the SO based on past data and is announced to the customers. The SO also announces its policy parameter $\gamma_h$ and its expectation of the renewable term $\bar\omega_h$. In addition, each customer knows its own consumption preference $g_{i,h}$.

\subsection{Consumer behavior models} \label{behavior_model_section}

Consumption behavior $\{l_{i,h}\}_{i = 1,\dots,N}$ determines the population's aggregate utility at time $h$,
\begin{equation}\label{aggregate_utility}
U_h(l_{i,h},l_{-i,h}):= \sum_{i} u_{i,h}(l_{i,h}, L_{h} ; g_{i,h},\omega_h).
\end{equation}
The net revenue of the SO is its revenue minus the cost
\begin{equation} \label{net_revenue}
NR_h(L_h;\omega_h) := p_h(L_h; \omega_h) L_h - C_h(L_h),
\end{equation}
where $C_h(L_h)$ is the cost of supplying $L_h$ Watts of power. When the generation cost per unit is constant, $C_h(L_h)$ is a linear function of $L_h$. More often, increasing the load $L_h$ results in increasing unit costs as the SO needs to dispatch power from more expensive sources. This results in superlinear cost functions with an approximate model being the quadratic form\begin{footnote}{{It is possible to add linear and constant cost terms to $C_h(L_h)$ and have all the results in this paper still hold. We exclude these terms to simplify notation.}}\end{footnote}
\begin{align} \label{provider_cost}
   C_h(L_h) =\frac{\kappa_h}{N} L_h^2
\end{align}
for a given time dependent constant $\kappa_h > 0$ normalized by number of consumers $N$. The cost in \eqref{provider_cost} has been experimentally validated for thermal generators \cite{Power_operation_control}, and it is otherwise widely accepted as a reasonable approximation \cite{SmartGrid_mechanism_design, Autonomous_DSM_SG, Atzeni_et_al}. The welfare of the overall system at time $h$ is the sum of the aggregate utility with the net revenue,
\begin{equation} \label{welfare}
W_h(l_{i,h},l_{-i,h}) := U_h(l_{i,h},l_{-i,h}) + NR_h(l_{i,h},l_{-i,h}).
\end{equation}

Consumer behavior can be selfish, altruistic or welfare-maximizing. User $i$ is \emph{selfish} when it wants to maximize its individual utility in \eqref{utility_i}. It is \emph{altruistic} when it considers the well-being of other users, that is, aims to  maximize $U_h$ in \eqref{aggregate_utility}. Finally, user $i$ might also consider the well-being of the whole system and aim to choose his consumption behavior to maximize the \emph{welfare} $W_h$ in \eqref{welfare} given its information.  We use the superscript $\Gamma \in $ \{S, U, W\} in $u_{i,h}^\Gamma(l_{i,h},l_{-i,h})$ to indicate that the consumer $i$ maximizes its selfish payoff S, aggregate utility U or the welfare W. All of these behavior models require strategic reasoning about the behavior of others which constitutes a Bayesian game. Bayesian games model interactions where users have incomplete information about the utility of others. Below we formalize a range of information exchange models.

\subsection{Information models} \label{information_exchange_section}

Consumption preference profile $\bbg_h$ is partially known by the individuals. Consumption decisions of individuals at time $h$ can provide valuable information about these consumption preferences. This information is of use to consumer $i$ in estimating consumption for the next time slot $h+1$ if the preferences of the users do not change in that time slot, that is, $\bbg_h = \bbg_{h+1}$. Otherwise, the information at time $h$ is not helpful in estimating the behaviors of others for time slot $h+1$ because we assume the change in the preference distribution to be independent. We let an uninterrupted sequence of time slots in which agents have the same consumption preference profile define a time zone. Formally, a time zone is defined as  $\ccalT = \{h \in \ccalH: \bbg_h = \bbg \wedge ((\bbg_{h-1}=\bbg) \vee (\bbg_{h+1} = \bbg)\vee \bbg)\}$ for a preference profile $\bbg := [g_{1}~\dots ~g_{N}]^T$ with prior probability density function $P_{\bbg}$ where $\wedge$ is `and' operator and $\vee$ is `or' operator. Next, we present a set of possible information exchange models within a time zone $\ccalT$. We use $I_{i,h}^\Omega$ to denote the set of information available to consumer $i$ at time slot $h \in \ccalT$ for the information exchange model $\Omega$. 

\begin{mylist}
\item[{\it Private.}] The information specific to consumers is the merest possible when it consists of the private  preference $g_{i}$, $I_{i,h}^P = \{g_{i}\}$ for $h\in \ccalT$.

\item[{\it Action-Sharing.}]  Power control schedulers are interconnected via a communication network represented by a graph $\ccalG(\ccalN,\ccalE)$ with its nodes representing the customers $\ccalN = \{1,\dots, N\}$ and edges belonging to the set $\ccalE$ indicating the possibility of communication. User $i$ observes consumption levels of his neighbors in the network $\ccalN_i := \{j\in \ccalN\! : (j,i)\in \ccalE\}$ after each time slot. The vector of $i$'s $d(i) := \# \ccalN_i$ neighbors is denoted by $[i_1, \dots, i_{d(i)}]$. Given the communication set-up, the information of user $i$ at time slot $h \in\ccalT $ contains its self-preference $g_{i}$ and the consumption of his neighbors up to time $h-1$, that is,  $I_{i,h}^{AS} = \{g_{i}, \{l_{\ccalN_i, t}\}_{t=1,\dots, h-1}\}$ where we define the actions of $i$'s neighbors at time $t$ by $l_{\ccalN_i, t}:= [l_{i_1,  t}, \dots, l_{i_{d(i)}, t}]$ and denote the starting time slot of $\ccalT$ with $t=1$. We assume that the power consumption schedulers keep the information received from neighbors private and know the network structure $\ccalG$. 

\item[{\it SO Broadcast.}] The SO collects all the individual consumption behavior at each time $h$ and broadcasts the total consumption to all the customers, that is, $I_{i,h}^B = \{g_{i}, L_{1:h-1}\}$.
\end{mylist}

When the time zone $\ccalT$ ends, we restart the information exchange process. The prediction of renewable source term $P_{\omega_h}$ is allowed to vary for $h\in\ccalT$. Behavior model, $\Gamma \in $ \{S, U, W\}, and the information exchange model, $\Omega \in $ \{P, AS, B\}, determine the consumption decisions of user $i$. In the following, we define the rational consumer behavior in Bayesian games within a time zone $\ccalT$ and then characterize the rational behavior for each behavior and information exchange model pair ($\Gamma,\Omega$). 

%% file: stage_game_solution.tex

User $i$'s load consumption at time $h \in \ccalT$ is determined by his \emph{strategy} $s_{i,h}$ that maps his information to a consumption level. This map depends on the \emph{belief} of $i$ which is a conditional probability on $\bbg$ and $\omega$ given its information, $P_{\bbg, \omega}(\cdot | I_{i,h}^\Omega)$. We use $E_{i,h}^\Omega[\cdot] := E_{\bbg, \omega}[\cdot|I_{i,h}^\Omega]$ to indicate conditional expectation with respect to its belief. While the model can account for the correlation between the random variables $\omega_h$ and $\bbg$, we assume that they are independent. In order to second-guess the consumption of other customers, user $i$ forms beliefs on preferences given the common prior $P_{\bbg}$ and its information $I_{i,h}^\Omega$. User $i$'s load consumption at time $h \in \ccalT$ is determined by its strategy which is a complete contingency plan that maps any possible local observation that it may have to its consumption; that is, $s_{i,h}\! : I_{i,h}^\Omega \mapsto \reals^+$ for any $I_{i,h}^\Omega$. In particular, for user $i$, its best response strategy is to maximize its expected utility given the strategies of other customers $\bbs_{-i,h}:= \{s_{j,h}\}_{j \neq i}$,
\begin{align} \label{best_response}
BR^\Gamma&( I_{i,h}^\Omega; \bbs_{-i,h}) = \arg \max_{l_{i,h}} E_{i,h}^\Omega\big[ u_{i,h}^\Gamma(l_{i,h},\bbs_{-i,h})\big]. 
\end{align}

Before we define the Bayesian Nash equilibrium (BNE) solution, we introduce the following lemma which characterizes the general form of the best response function for all the behavior models $\Gamma \in $ \{S, U, W\}.

\begin{lemma} \label{best_response_lemma}
The best response strategy of $i$ to the strategies of others $\bbs_{-i,h}$ has the following general form for any behavior model $\Gamma \in\rm{ \{S, U, W\} }$
\begin{align} \label{best_response_i}
BR^\Gamma&( I_{i,h}^\Omega; \bbs_{-i,h}) =\frac{g_{i} - \mu_h^\Gamma \bar\omega_h- \lambda_h^\Gamma \sum_{j \neq i} E_{i,h}^\Omega[s_{j,h}] }{2(\tau_h^\Gamma + \alpha_h)}
\end{align} 
where $\lambda_h^\Gamma, \mu_h^\Gamma, \tau_h^\Gamma$ are constants that take values based on the behavior model $\Gamma$. If $\Gamma = \rm{S}$ then $\lambda_h^S = \mu_h^S =\tau_h^S = \gamma_h/N$. If $\Gamma = \rm{U}$ then $\lambda_h^{U} = 2 \gamma_h/N$, $\mu_h^{U} = \tau_h^{U} = \gamma_h/N$. If $\Gamma = \rm{W}$ then $\lambda_h^W= 2 \kappa_h/N$, $\mu_h^W = 0$, $\tau_h^W = \kappa_h/N$.
\end{lemma}

The proof follows by taking the derivative of the corresponding utility with respect $i$'s consumption $l_{i,h}$, equating to zero and solving the equality for $l_{i,h}$. Note that when $\bar\omega_h = 0$ and $\gamma_h = \kappa_h$, the altruistic users have the same best response function as the welfare-maximizers. A BNE strategy profile for the game $\Gamma$ is a strategy in which each user maximizes its expected utility $u_{i,h}^\Gamma$ with respect to its own belief given that other users also maximize their expected utility \cite[Ch.6]{Fudenberg_Tirole_1991}.

\begin{definition} A BNE strategy $\bbs^\Gamma := \{s^\Gamma_{i,h}\}_{i \in\ccalN, h\in \ccalT}$ for the consumer behavior model $\Gamma \in \rm{\{S,U,W\}}$ is such that for all $i \in \ccalN$, $h \in \ccalT$, and $\{I_{i,h}^\Omega\}_{i\in \ccalN, h\in \ccalT}$, 
\begin{align} \label{BNE}
E_{i,h}^\Omega\big[ u_{i,h}^\Gamma&(s_{i,h}^\Gamma, \bbs_{-i,h}^\Gamma)\big]  \geq  E_{i,h}^\Omega\big[ u_{i,h}^\Gamma(s_{i,h}, \bbs_{-i,h}^\Gamma) \big].
\end{align}
for any $s_{i,h}: I_{i,h}^\Omega \mapsto \reals^+$.
\end{definition}
A BNE strategy \eqref{BNE} is computed using beliefs formed according to Bayes' rule. Note that the BNE strategy profile is defined for all time slots. No user at any given time slot within $\ccalT$ has a profitable deviation to another strategy. 

In \eqref{BNE}, consumers estimate consumption decisions of others to respond optimally. Equivalently, a BNE strategy is one in which users play best response strategy given their individual beliefs as per \eqref{best_response} to best response strategies of other users -- see \cite{Eksin_et_al_2013,Ho_Chu} for similar equilibrium concepts. As a result, the BNE strategy is defined by the following fixed point equations:
\begin{align} \label{BNE_fixed_point}
s_{i,h}^\Gamma(I_{i,h}^\Omega) = BR(I_{i,h}^\Omega; \bbs_{-i,h}^\Gamma) 
\end{align}
for all $i \in \ccalN$, $h \in \ccalT$, and $I_{i,h}^\Omega$. We denote $i$'s realized load consumption from the equilibrium strategy $s_{i,h}^\Gamma$ and information $I_{i,h}^\Omega$ with $l_{i,h}^\Gamma := s_{i,h}^\Gamma(I_{i,h}^\Omega)$. Using the definition in \eqref{BNE_fixed_point}, we characterize the unique linear BNE strategy in the next section for any information exchange and consumer behavior model. 

%% file: communication.tex

It suffices for customer $i$ to estimate the self-preference profile $\bbg$ in order to estimate consumption of other users  \cite{Eksin_et_al_2013}. We define the self-preference profile augmented with mean $\bar{g}$ as  $\tilde{\bbg}:=[\bbg^T, \bar{g}]^T$. The mean and error covariance matrix of $i$'s belief at time $h$ are denoted by $E_{i,h}^\Omega[\tilde{\bbg}]$ and $\bbM^i_{\tilde{\bbg} \tilde{\bbg}}(h):=E[(\tilde{\bbg}-E_{i,h}^\Omega[\tilde{\bbg}])(\tilde{\bbg}-E_{i,h}^\Omega[\tilde{\bbg}])^T]$, respectively. The next result shows that there exists a unique BNE strategy that is a linear weighting of the mean estimate of $\tilde{\bbg}$ for any information model $\Omega$. Furthermore, the weights of the linear strategy are obtained by solving a set of linear equations specific to the behavior model $\Gamma$. 
\begin{proposition}\label{BQNG_theorem} 
Consider the Bayesian game defined by the payoff $u_{i,h}^\Gamma$ for $\Gamma \in {\rm \{S, U, W\}}$. Let the information $I_{i,h}^\Omega$ of customer $i$ at time $h \in \ccalT$ be defined by the information exchange model $\Omega \in  {\rm \{P, AS, B\}}$. Given the normal prior on the self-preference profile $\bbg$, user $i$'s  mean estimate of the preference profile at time $h \in \ccalT$ can be written as a linear combination of $\tilde{\bbg}$. That is, $E_{i,h}^\Omega[\tilde{\bbg}] = \bbT_{i,h}^\Omega \tilde{\bbg}$ where $\bbT_{i,h}^\Omega\in \reals^{N+1 \times N+1}$ for all $h \in \ccalT$, and the unique equilibrium strategy for $i$ is linear in its estimate of the augmented self-preference profile, 
\begin{equation}\label{linear_equilibrium_strategy}
s_{i,h}^\Gamma(I_{i,h}^\Omega) = \bbv_{i,h}^{T} E_{i,h}^\Omega[\tilde{\bbg}] + r_{i,h}
\end{equation}
where $\bbv_{i,h} \in \reals^{N+1 \times 1}$  and $r_{i,h}\in \reals$ are the strategy coefficients. The strategy coefficients are calculated by solving the following set of equations for the consumer behavior models $\Gamma \in {\rm \{S, U, W\}}$
\begin{equation}
\bbv_{i,h}^{T} \bbT_{i,h}^{\Omega T}  + \rho_h^\Gamma \lambda_h^\Gamma \sum_{j \in \ccalN\setminus i} \bbv_{j,h} \bbT_{i,h}^{\Omega T}  \bbT_{j,h}^{\Omega T} = \rho_h^\Gamma \bbe_i , \quad \forall \; i \in \ccalN , \label{first_equation_set}
\end{equation}
and
\begin{equation}
r_{i,h} + \rho_h^\Gamma \lambda_h^\Gamma \sum_{j \in \ccalN\setminus i} r_{j,h}^\Gamma = -\rho_h^\Gamma \mu_h^\Gamma \bar\omega_h , \quad \forall i \in \ccalN , \label{second_equation_set}
\end{equation}
where $\lambda_h^\Gamma, \mu_h^\Gamma, \tau_h^\Gamma$ are as defined in Lemma \ref{best_response_lemma} for $\Gamma \in {\rm \{S,U,W\}}$, 
$\rho_h^\Gamma = (2(\tau_h^\Gamma+\alpha_h))^{-1}$ and $\bbe_i \in \reals^{N+1 \times 1}$ is the unit vector.
\end{proposition}
\begin{myproof}\begin{footnote}{The proof is adopted from Proposition \ref{BQNG_theorem} in \cite{Eksin_et_al_2013}.}\end{footnote}Our plan is to propose a linear strategy and use the general form of the best response function \eqref{best_response_i} in the fixed point equations \eqref{BNE_fixed_point} to obtain the set of linear equations. We prove by induction. Assume that users have linear estimates at time $h $, $E_{i,h}^\Omega[\tilde{\bbg} ] = \bbT_{i,h}^\Omega \tilde{\bbg}$ for all $i \in \ccalN$. We propose that users follow a strategy that is linear in their mean estimate as in \eqref{linear_equilibrium_strategy}. Using the fixed point definition of BNE strategy in \eqref{BNE_fixed_point}, we have
\begin{equation}\label{BNE_1}
\bbv_{i,h}^T E_{i,h}^\Omega[\tilde{\bbg}] + r_{i,h} =  \frac{g_{i} - \mu_h^\Gamma \bar\omega_h- \lambda_h^\Gamma \sum_{j \neq i} E_{i,h}^\Omega[\bbv_{j,h}^T E_{j,h}^\Omega[\tilde{\bbg}] + r_{j,h}]}{2(\tau_h^\Gamma + \alpha_h)}
\end{equation}
for all $i \in \ccalN$ from Lemma \ref{best_response_lemma}. The summation above includes user $i$'s expectation of user $j$'s expectation of the augmented preferences. By the induction hypothesis, we write this term as
\begin{equation}
E[E[\tilde{\bbg} | I_{j,h}^\Omega]  | I_{i,h}^\Omega] = \bbT_{j,h}^\Omega \bbT_{i,h}^\Omega \tilde{\bbg} .
\end{equation}
Substituting the above equation for the corresponding terms in \eqref{BNE_1} and using the induction hypothesis for the expectation term on the left-hand side yields the set of equations
\begin{equation}
\bbv_{i,h}^T \bbT_{i,h}^\Omega \tilde{\bbg} + r_{i,h} =  \frac{g_{i}-  \mu_h^\Gamma \bar\omega_h - \lambda_h^\Gamma \sum_{j \neq i} \bbv_{j,h}^T  \bbT_{j,h}^\Omega \bbT_{i,h}^\Omega\tilde{\bbg}  + r_{j,h}}{2(\tau_h^\Gamma + \alpha_h)}.
\end{equation}
We equate the terms that multiply $\tilde{\bbg}$ and the constants to obtain the set of equations in \eqref{first_equation_set} and \eqref{second_equation_set}, respectively. 

Since user consumption is based on its BNE strategy at time $h$, it is linear in its estimate of the preferences; i.e., $l_{j,h}^\Gamma = \bbv_{i,h}^T \bbT_{i,h}^\Omega\tilde{\bbg}  + r_{i,h}$ for all $j \in \ccalN$. We can then express the observations of user $i$ as a linear combination of $\tilde{\bbg}$ by defining the observation matrix $\bbH_{i,h}^{\Omega}$ for any information exchange model $\Omega \in {\rm \{P, AS, B\}}$. For the private information model, the observation matrix is zero, i.e., $\bbH_{i,h}^{{\rm P}} = \bbzero$ for any $h \in \ccalT$. For the action-sharing information model, the observations of consumer $i$ can be written using the observation matrix $\bbH_{i,h}^{{\rm AS}}\in \reals^{d(i) \times N+1}$
\begin{equation}\label{observation_matrix_share}
\bbH_{i,h}^{{\rm AS}}:= [\bbv_{j_{i1},h}^T \bbT_{j_{i1},h}^{{\rm AS}}; \ldots; \bbv_{j_{i d(i)},t}^T \bbT_{j_{i d(i)},h}^{{\rm AS}}]^T 
\end{equation} 
and the vector $\bbr_{\ccalN_{i},h}:= [r_{j_{i1},h}; \ldots; r_{j_{i d(i)},h}]$, as 
$
   l_{\ccalN_i, h}^{{\rm AS}} = \bbH_{i,h}^{{\rm AS}} \tilde\bbg + \bbr_{\ccalN_i,h}.
$
Finally, when the SO broadcasts the total consumption $L_h^{\rm B}$, the observation matrix is a vector 
\begin{align} \label{observation_matrix_broadcast}
\bbH_{i,h}^{{\rm B}}  = \sum_{j =1}^N (\bbv_{j,h}^T \bbT_{j,h}^{\rm B})^T ,
\end{align}
and the total consumption can be written as 
$   L_{h}^{\rm B} = \bbH_{i,h}^{{\rm B}} \tilde\bbg + \sum_{j=1}^N \bbr_{j,h}.
$
Because the prior distribution on the preferences are Gaussian, the observations of user $i$ are Gaussian for all information exchange models $\Omega \in {\rm \{P, AS, B\}}$. As a result, we can use a Kalman filter with gain matrix 
\begin{IEEEeqnarray}{rCl}
K^i_{\tilde\bbg}(h) & :=  & \bbM^i_{\tilde\bbg \tilde\bbg} (h) \bbH_{i,h}^\Omega\big(\bbH_{i,h}^{\Omega T}  \bbM^i_{\tilde\bbg \tilde\bbg} (h) \bbH_{i,h}^\Omega \big)^{-1} \label{eqn_lmmse_gain_x}
\end{IEEEeqnarray}
to propagate mean beliefs in the following way:
\begin{align}
E\left[\tilde\bbg\given I_{i,h+1}^\Omega\right] &= E \left[\tilde\bbg \given I_{i,h}^\Omega\right] +  K^i_{\tilde\bbg}(h) \left(\bbH_{i,h}^{\Omega T} \tilde\bbg-  \bbH_{i,h}^{\Omega T} \bbT_{i,h}^\Omega \tilde\bbg\right).\label{LMMSE_updates_mean_10}
\end{align}
We use the induction hypothesis  $E[\tilde{\bbg} | I_{i,h}^\Omega] = \bbT_{i,h}^\Omega \tilde{\bbg}$ for the first term on the right hand side of \eqref{LMMSE_updates_mean_10} and rearrange terms to get
\begin{align}
E\left[\tilde\bbg\given I_{i,h+1}^\Omega\right] &= \big( \bbT_{i,h}^\Omega+  K^i_{\tilde\bbg}(h) \left(\bbH_{i,h}^{\Omega T} -  \bbH_{i,h}^{\Omega T} \bbT_{i,h}^\Omega \right) \big)\tilde\bbg. \label{linear_weights_update}
\end{align}
Note that the mean estimate at time $h+1$ is a linear combination of $\tilde\bbg$. Specifically, we can express the linear weights of the mean estimate at time slot $h+1 $ as 
\begin{align}
\bbT_{i,h+1}^\Omega &= \bbT_{i,h}^\Omega +  K^i_{\tilde\bbg}(h) \Big(\bbH_{i,h}^{\Omega T} -  \bbH_{i,h}^{\Omega T} \bbT_{i,h}^\Omega\Big) \label{weights_recursion_x}
\end{align} 
where the mean estimate is $E\left[\tilde\bbg\given I_{i,h+1}^\Omega\right] = \bbT_{i,h+1}^\Omega \tilde\bbg$, completing the induction argument. Similarly, the updates for error covariance matrices follow standard Kalman updates \cite[Ch. 12]{Kay}
\begin{align}
\bbM^i_{\tilde\bbg \tilde\bbg} (h+1) =& \bbM^i_{\tilde\bbg \tilde\bbg} (h)- K^i_{\tilde\bbg}(h) \bbH_{i,h}^{\Omega T} \bbM^i_{\tilde\bbg \tilde\bbg} (h).\label{LMMSE_updates_covariance}
\end{align}
At the starting time slot $h =1$, we have $E[g_{j} \given g_{i}] = (1- \sigma_{ij}/\sigma_{ii})\bar{g}+ (\sigma_{ij}/\sigma_{ii}) g_{i}$. Hence the induction assumption is true initially and $E_{i,1}^\Omega[\tilde\bbg] = E[\tilde\bbg \given g_{i}] = \bbT_{i 1}^\Omega \tilde\bbg$ for all $\Omega \in {\rm \{P, AS, B\}}$.

Since the stage game has the same pay-off structure and the information is Gaussian, it suffices to show uniqueness for the stage game. The uniqueness of the stage game is proven in Proposition 1 in \cite{Eksin_et_al_15}. See also Proposition 2.1 in \cite{Vives_2008}. 
\end{myproof}

Proposition \ref{BQNG_theorem} presents how BNE consumption strategies are computed at each time slot. Accordingly, the scheduler repeatedly determines its consumption strategy given consumption behavior model $\Gamma$ and available information, receives information based on the information exchange model $\Omega$ at the end of the time slot, and propagates its beliefs on self-preference profile to be used in the next time slot. 
For each consumption behavior $\Gamma \in $ \{S, U, W\} the user solves a different set of equations in \eqref{first_equation_set}-\eqref{second_equation_set} derived from the fixed point equations of the BNE \eqref{BNE_fixed_point}. For \emph{Private} information exchange model, users do not receive any new information within the horizon hence their mean estimate of $\tilde{\bbg}$ do not change, that is, $\bbT_{i,h}^P = \bbT_{i, 1}^P$ for $h \in \ccalT$, which implies the set of equations \eqref{first_equation_set}-\eqref{second_equation_set} need to be solved only once at the beginning to determine the strategy for the whole time horizon. For \emph{Action-Sharing} information exchange model, upon observing actions of its neighbors, user $i$ has new relevant information about the preference profile which it can use to better predict the total consumption in future steps. Similarly in \emph{SO Broadcast} model, each user receives the total consumption at each time which is useful in estimating total consumption in the following time slot. 

The Bayesian belief propagation for Gaussian prior beliefs corresponds to Kalman filter updates at each step for any information exchange model. In particular, beliefs remain Gaussian and the mean estimates are linear combinations of private signals at all times for any information exchange model. In order to compute the BNE strategy, it does not suffice for scheduler $i$ to form beliefs on the preference $\tilde{\bbg}$. It also needs to keep track of beliefs of others. Knowing the estimate of all the other schedulers is not possible for $i$. However, this is not required to compute an estimate of other schedulers' estimates. It is only required that user $i$ knows how other schedulers compute their mean estimates which implies knowing the estimation weights $\bbT_{j,h}^\Omega$. Even though scheduler $i$ does not know $P_{\tilde\bbg}(\tilde\bbg | I_{j,h}^\Omega)$, it can keep track of $\bbT_{j,h}^\Omega$ via the weight recursion equation in \eqref{weights_recursion_x}, which can be computed using public information. Note that $i$ cannot compute self-mean estimate of preferences, $E \left[\tilde\bbg \given I_{i,h}^\Omega\right]$, via multiplying $\bbT_{i,h}^\Omega$ by $\tilde\bbg$ since this computation would require knowledge of $\tilde\bbg$. Instead, user $i$ computes its mean estimate by a Kalman filter. We detail the local computations of a scheduler in Algorithm \ref{alg1}.

In Algorithm \ref{alg1}, we provide a local algorithm for user $i$ to compute its consumption level and propagate its belief given a behavior model $\Gamma\in$ \{S, U, W\} and the information exchange model $\Omega =$ AS. We point to modifications specific to the other information exchange models here in our explanation. User $i$ initializes its belief on $\tilde{\bbg}$ at the beginning of the time zone $\ccalT$ according to the preference distribution in \eqref{preference_distribution}. It also determines the estimation weights $\bbT_{j,1}^\Omega$ and error covariance matrix $\bbM^j_{\tilde\bbg \tilde\bbg} (1)$  at the beginning for $j \in \ccalN$. Note that user $i$ does not need any local information from other users in this initialization. Using the estimation weights $\{\bbT_{j,1}^\Omega\}_{j \in \ccalN}$, it can locally  construct the equations in \eqref{first_equation_set} and \eqref{second_equation_set}, and  solve for the strategy coefficients $\{\bbv_{j,h}, r_{j,h}\}_{j \in \ccalN}$. In Step 2, $i$ consumes the amount based on its local estimate of the augmented self-preferences -- see \eqref{linear_equilibrium_strategy}. 

Once the consumption occurs, the information becomes available according to the information exchange model $\Omega$. At this point, if the upcoming time slot $h+1$ has the same prior preference distribution \eqref{preference_distribution} as $h$, that is, if $h+1 \in \ccalT$, $i$ propagates its belief on the self-preference profile given the new information. The propagation of beliefs starts by computing observation matrices of \emph{all} the users in Step 3 based on the information exchange model $\Omega$. When the model is \emph{action-sharing}, $\Omega =$ AS, each observed action $\{l_{j,h}^{\rm AS}\}_{j \in \ccalN_i}$ is a linear combination of $\tilde\bbg$ with the observation matrix $\bbH_{j,h}^{\rm AS}$ computed by  \eqref{observation_matrix_share}. If the model is \emph{broadcast}, $\Omega =$ B, the observation matrix is a vector computed by \eqref{observation_matrix_broadcast}. If the model is \emph{private}, $\Omega =$ P, there is no new information available hence scheduler $i$ goes back to Step 2 with the same strategy coefficients. Next, $i$ uses these observation matrices in computing the gain matrices in Step 4 of all the users. In Step 5, $i$ propagates the estimation weights $\bbT_{j,h +1 }^{\rm B}$ and error covariance matrix $\bbM^j_{\tilde\bbg \tilde\bbg} (h+1)$. Note that in Steps 3-5 user $i$ does a full network simulation in which it emulates the Kalman filter estimates of everyone using public information, that is, estimation weights $\{\bbT_{j,h}^\Omega\}_{j \in \ccalN}$, strategy coefficients $\{\bbv_{j,h}\}_{j \in \ccalN}$ and network topology $\ccalG$. Finally in Step 6, $i$ propagates its \emph{own} mean estimate $E\left[\tilde\bbg\given I_{i,h+1}^\Omega\right]$ by using its own local observation, which is $l_{\ccalN_i,h}^{\Gamma}$ for $\Omega = $ AS or $L_h^\Gamma$ for $\Omega=$ B.

%
\begin{algorithm} [t]             
\caption{Sequential Game Filter for $\Omega = AS$ at User $i$}          
\label{alg1}                           
\begin{algorithmic}                    
 \REQUIRE Consumer behavior model $\Gamma \in$ \{S, U, W\}.
 \REQUIRE Posterior distribution on $\tilde\bbg$ at time slot $h = 1$ and $\{\bbT_{j,1}^\Omega, \bbM^j_{\tilde\bbg \tilde\bbg} (1)\}_{j\in \ccalN}$ according to \eqref{preference_distribution}. 
\WHILE{$\bbg_h =\bbg$}
\STATE[1]  \emph{Equilibrium $\Gamma$:}  Solve $\{\bbv_{j,h}, r_{j,h}\}_{j \in \ccalN}$ using \eqref{first_equation_set}-\eqref{second_equation_set}. \\
\STATE[2]  \emph{Play}: Compute $s_{i,h}^\Gamma(I_{i,h}^\Omega) = \bbv_{i,h}^T E[\tilde\bbg \given I_{i,h}^\Omega] +r_{i,h}$. 
\STATE[3] \emph{Construct observation matrix $\{\bbH_{j,h}^\Omega\}_{j \in \ccalN}$}: Use  \eqref{observation_matrix_share}. \\
\STATE[4] \emph{Gain matrices:} Compute $\{\bbK^j_{\tilde\bbg}(h)\}_{j \in \ccalN}$  
\begin{align}
\bbK^j_{\tilde\bbg}(h)  :=   \bbM^j_{\tilde\bbg \tilde\bbg} (h) \bbH_{j,h}^\Omega\big(\bbH_{j,h}^{\Omega T}  \bbM^j_{\tilde\bbg \tilde\bbg} (h) \bbH_{j,h}^{\Omega} \big)^{-1} \nonumber
\end{align}
 
\STATE[5] \emph{Estimation weights:} Update $\{\bbT_{j,h+1}, \bbM^j_{\tilde\bbg \tilde\bbg} (h+1)\}_{j\in \ccalN}$  
\begin{align}
\bbT_{j,h+1} &= \bbT_{j,h}^\Omega + \bbK^j_{\tilde\bbg}(h) \Big(\bbH_{j,h}^{\Omega T} -  \bbH_{j,h}^{\Omega T} \bbT_{j,h}^\Omega \Big)\nonumber 
\end{align} 
\begin{align}
\bbM^j_{\tilde\bbg \tilde\bbg} (h+1) =& \bbM^j_{\tilde\bbg \tilde\bbg} (h)- \bbK^j_{\tilde\bbg}(h) \bbH_{j,h}^{\Omega T} \bbM^j_{\tilde\bbg \tilde\bbg} (h).\nonumber 
\end{align}
 
\STATE[6] \emph{Bayesian estimates:}  Calculate $E[\tilde\bbg \given I_{i,h+1}^\Omega]$
\begin{align}
E[\tilde\bbg & \hspace{-1pt}\given \hspace{-1pt} I_{i,h+1}^{\Omega}] \hspace{-1pt}= \hspace{-1pt}
E \left[\tilde\bbg \given I_{i,h}^{\Omega}\right]  \hspace{-1pt}+ \hspace{-1pt}  \bbK^i_{\tilde\bbg}(h)  \big(l_{\ccalN_i,h}^\Gamma   \hspace{-1pt}- \hspace{-1pt}E[l_{\ccalN_i,h}^\Gamma\given I_{i,h}^{\Omega}]\big).  \nonumber
\end{align}
\ENDWHILE
\end{algorithmic}
\end{algorithm}
\subsection{Private and complete information games} \label{private_full_information_section}

In Step 2 of Algorithm \ref{alg1} the user solves a set of $N^2$ linear equations.  This computation can be avoided in situations where the information of each consumer remains the same. The information is static in the two extreme cases. The first extreme case is when the information exchange model is private ($\Omega = $ P). For the private information case, there exists a closed-form solution to the set of equations in \eqref{first_equation_set}-\eqref{second_equation_set} that is symmetric when the preference correlation is homogeneous. We state this result in the following.
\begin{proposition}\label{cor_BNE_characterization_private}
Consider the Bayesian game defined by the payoff $u_{i,h}^\Gamma$ for $\Gamma \in {\rm \{S, U, W\}}$ and the the private information exchange model $\Omega =  {\rm P}$. Assume the preferences of users are $\sigma$-correlated at time $h$, that is, the off-diagonal elements of $\Sigma$ are the same $\sigma_{ij} = \sigma \in [0,1]$ for all $i=1,\dots,N$ and $j \in\{1,\dots,N\}\setminus i$ and its diagonal elements equal to 1. Then, the unique BNE strategy of user $i$ is
linear in $\bar{\omega}_h$, $\bar{g}_h$, $g_{i,h}$ for $\Gamma \in {\rm \{S, U, W\}}$ such that
\begin{equation} \label{eq_explicit_private}
s_{i,h}^{\Gamma}(I_{i,h}^{\rm P}) = a_h^\Gamma (g_{i,h}-\bar{g}_h) + b_h^\Gamma (\bar{g}_h - \bar{\omega}_h \mu_h^\Gamma)
\end{equation}
where we define constants $a_h^\Gamma = \rho_h^\Gamma (1+\lambda_h^\Gamma \rho_h^\Gamma \sigma (N-1))^{-1}$, $b_h^{\Gamma} = \rho_h^\Gamma (1+\lambda_h^\Gamma \rho^\Gamma(N-1))^{-1}$, and $\rho_h^{\Gamma} = (2(\tau_h^\Gamma+ \alpha_h))^{-1}$.
\end{proposition}
\begin{myproof}
See Proposition 1 and 2 in \cite{Eksin_et_al_15} for the proof of the selfish case ($\Gamma = $ S). Proofs for the other cases follow the same steps where we consider the best response function in \eqref{best_response_i} of the corresponding behavior model. 
\end{myproof}

Second extreme case of static information is when all the users have complete information. For the game we consider, for each customer, his private preference and the cumulative realized preference $\{g_{i}, \sum_{j}g_{j,h}\}$ form a sufficient statistic of the realized preferences $\bbg$ for the homogeneously correlated preference games $\Gamma \in $ \{S, U, W\} -- see \cite{Vives_2011}. Next we provide an explicit characterization of BNE strategies when users have complete information. 
\begin{proposition}\label{cor_BNE_characterization_complete}
Consider the Bayesian game defined by the payoff $u_{i,h}^\Gamma$ for $\Gamma \in {\rm \{S, U, W\}}$. Let the information of customer $i$ at time $h \in \ccalT$ be $I_{i,h} = \{g_{i}, \sum_{j}g_{j,h}\}$. Assume the preferences of users are $\sigma$-correlated at time $h$ as defined in Proposition \ref{cor_BNE_characterization_private}. Then, the unique BNE strategy of user $i$ is linear in $\bar{\omega}_h$, $ \sum_{j}g_{j,h}$, $g_{i,h}$ for all $h \in \ccalH$ such that
\begin{equation}\label{eq_explicit_broadcast}
s_{i,h}^{\Gamma}(I_{i,h}^B) =  a_h^\Gamma (g_{i,h}- \frac{1}{N}\sum_{j=1}^Ng_{j,h} ) + b_h^\Gamma (\frac{1}{N}\sum_{j=1}^N g_{j,h} - \bar{\omega}_h \mu_h^\Gamma) 
\end{equation}
where we define constants $a_h^\Gamma = \rho_h^\Gamma (1-\lambda_h^\Gamma \rho_h^\Gamma)^{-1}$, $b_h^{\Gamma} = \rho_h^\Gamma   (1+\lambda_h^\Gamma \rho_h^\Gamma (N-1))^{-1}$ and $\rho_h^{\Gamma} = (2(\tau_h^\Gamma+ \alpha_h))^{-1}$.
\end{proposition}
\begin{myproof}
Proof follows along the similar lines of the proof of Proposition \ref{cor_BNE_characterization_private}.  
\end{myproof}

Complete information is achieved when the SO broadcasts total consumption $L_h$ and the preference correlation is homogeneous. That is, the total consumption $L_h$ conveys the cumulative realized preference $\sum_{j}g_{j}$ when the preferences are homogeneously correlated. Thus, in the broadcast information exchange model, $\Omega = B$, consumers play a private information game in the first time slot, and they have complete information from the second slot onwards. 

These characterizations allow for computation of BNE behavior by putting the available information into the linear strategy functions instead of following the steps of Algorithm \ref{alg1}. In the following sections, these characterizations allow for analytical comparison of performance measures with respect to different behavior and information exchange models.

%% file: simulation.tex

\section{Effects of behavior and information exchange models on demand}\label{sec_effects_demand}

We first focus on the effect private and complete information models have on demand. In both of these cases information is static. Hence it suffices to focus on demand at a given time $h$ for comparison. In the following, we drop the $h$ sub-index from the notation until we consider the action-sharing information exchange model. We define demand for a consumer behavior model $\Gamma\in {\rm \{S, U, W\}}$ as $L^\Gamma= \sum_{j=1}^N s_j^\Gamma$.

For the private and complete information cases, when the preferences are $\sigma$-correlated, the expected consumption of individuals are the same as per \eqref{eq_explicit_private} and \eqref{eq_explicit_broadcast}. Moreover, both the private and complete information models have the same expected total consumption, 
\begin{equation} \label{expected_demand}
E[L^\Gamma/N]=E[s_{i}^\Gamma(I_{i}^P)] = E[s_{i}^\Gamma(I_{i}^B)] = b^\Gamma (\bar{g} - \bar\omega \mu^\Gamma)
\end{equation}
by the fact that $b^\Gamma$ has the same value in Propositions \ref{cor_BNE_characterization_private} and \ref{cor_BNE_characterization_complete}. This means that allowing users to exchange information does not alter the demand forecasts of the operator. We can also read from the expected demand above that a welfare-maximizing user is not impacted by the changes in $\bar\omega$ because $\mu^W=0$. On the other hand, consumption of selfish and altruistic users decreases with slope $b^\Gamma \mu^\Gamma$ as the renewable term $\bar\omega$ increases. 

Given the explicit expected demand above, we compare the effects of different consumer behavior models. Throughout the analysis, we assume the preferences are $\sigma$-correlated where the diagonal elements of the covariance matrix are equal to 1 and off-diagonal elements are equal to $0 \leq\sigma\leq 1$.
\begin{corollary} \label{corollary_expected_consumption}
For large number of consumers, $N$, we have
\begin{enumerate}
\item $E[s_{i}^S]/E[s_{i}^U]= 2(\gamma + \alpha)/(\gamma + 2\alpha)>1$. 
\item If $\bar\omega = 0$ then $E[s_{i}^W]/E[s_{i}^U]= (\gamma + \alpha)/(\kappa + \alpha)$, which is greater than 1 for $\gamma >\kappa$.  
\item If $\bar\omega = 0$ then $E[s_{i}^S]/E[s_{i}^W]= 2(\kappa + \alpha)/(\gamma + 2\alpha)$, which is greater than 1 for $\gamma <2 \kappa$.  
\end{enumerate}
\end{corollary}
\begin{myproof}
When $N$ is large, we can approximate the $b^\Gamma$ behavior constants given in Proposition \ref{cor_BNE_characterization_private} as $b^S \approx (\gamma + 2 \alpha)^{-1}$, $b^U \approx (2(\gamma +  \alpha))^{-1}$ and $b^W \approx (2(\kappa +  \alpha))^{-1}$. The comparisons follow from \eqref{expected_demand}.
\end{myproof}

The above result says that the expected consumption of a selfish individual is higher than an altruistic user. Furthermore, the ordering of expected consumptions follow $E[s_i^S] > E [s_i^W] > E[s_i^U]$ when $\kappa<\gamma< 2 \kappa$. This range of price parameter $\gamma$ is of importance because the realized rate of return of the operator is approximately $\gamma/\kappa$. Hence, the operator is likely to select the pricing policy $\gamma$ from this range \cite{Eksin_et_al_15}. 

Next, we consider demand variance as an indicator of uncertainty that an operator has in its demand forecast.  We first focus on the private information model. 

\begin{corollary}
Consider the $\Omega =$ P information model with $\sigma$-correlated preferences.  We have the variance of normalized demand $L^\Gamma/N = \sum_{j=1}^N s_i^\Gamma(I_i^P)/N$ as follows:
\begin{equation} \label{eq_variance_demand}
Var(L^\Gamma/N) =  \frac{1+  (N-1)\sigma}{N}(a^\Gamma)^2 ,
\end{equation}
where $a^\Gamma$ is as defined in Proposition \ref{cor_BNE_characterization_private}.
\end{corollary}

\begin{myproof}
From the definition of the variance of normalized demand, we have
\begin{align}
Var(L^\Gamma/N) &:= \bigg(E[(\sum_{j=1}^Ns_i^\Gamma(I_i^P))^2] - E[(\sum_{j=1}^Ns_i^\Gamma(I_i^P))]^2\bigg)/N^2 \nonumber\\
&= E\bigg[\bigg(\sum_{j=1}^N a^\Gamma (g_i - \bar g)\bigg)^2\bigg]/N^2. \nonumber
\end{align}
In the second equality we use \eqref{eq_explicit_private} to cancel out the squared mean of consumption. The result follows from the above equation. 
\end{myproof}

Given the explicit representation of variance, we can make the following comparison about the effects of consumption behavior models on demand uncertainty. 
\begin{corollary} \label{cor_variance_demand_private}
Consider the $\Omega =$ P information model with $\sigma$-correlated preferences. For large number of consumers, $N$,  we have:
\begin{enumerate}
\item $Var(L^\Gamma/N) \approx \sigma (N\lambda^\Gamma \sigma + 2\alpha)^{-2}$, where $\lambda^\Gamma$ is as defined in Lemma \ref{best_response_lemma} for $\Gamma = \{{\rm S, U, W}\}$.
\item 
\begin{equation} \label{variance_change}
\frac{\partial Var(L^\Gamma/N)}{\partial \sigma} = \frac{2 \alpha - N \lambda^\Gamma \sigma }{(N \lambda^\Gamma \sigma + 2\alpha)^3}.
\end{equation}
\item If $\sigma = 0$, then the variance of normalized demand is the same for all behavior models $\Gamma = \{{\rm S, U, W}\}$. 
\item If $\sigma \in (0,1]$, then $1<Var(L^S/N)/Var(L^U/N)<4$.
\item If $\sigma \in (0,1]$, then $Var(L^U/N)/Var(L^W/N)=\frac{(\alpha + \kappa \sigma)^2}{(\alpha + \gamma \sigma)^2}$ which is less than 1 for $\gamma > \kappa$. 
\end{enumerate}
\end{corollary}
\begin{myproof}
When $N$ is large, we have the constant $a^\Gamma \approx (N \lambda^\Gamma \sigma + 2 \alpha)^{-1}$ from Proposition \ref{cor_BNE_characterization_private} where $\lambda^\Gamma$ is as defined in  Lemma \ref{best_response_lemma}. The comparisons follow from \eqref{eq_variance_demand}.
\end{myproof}

Equation \eqref{variance_change} shows that, with increasing $\sigma$, the demand variance grows as long as $N\lambda^\Gamma \sigma < 2 \alpha$ while it decreases with increasing $\sigma$ when  $N\lambda^\Gamma \sigma >2 \alpha$. Furthermore, the change in variance slows as $\sigma$ becomes larger.  From the fourth observation we see that selfish behavior achieves a higher demand variance in comparison to altruistic behavior. Next we focus on the demand variance in the complete information case that is reached after the operator broadcasts the total consumption information. 

\begin{figure}[!t]
\centering
\begin{tabular}{cc} 
\includegraphics[width=0.47\linewidth]{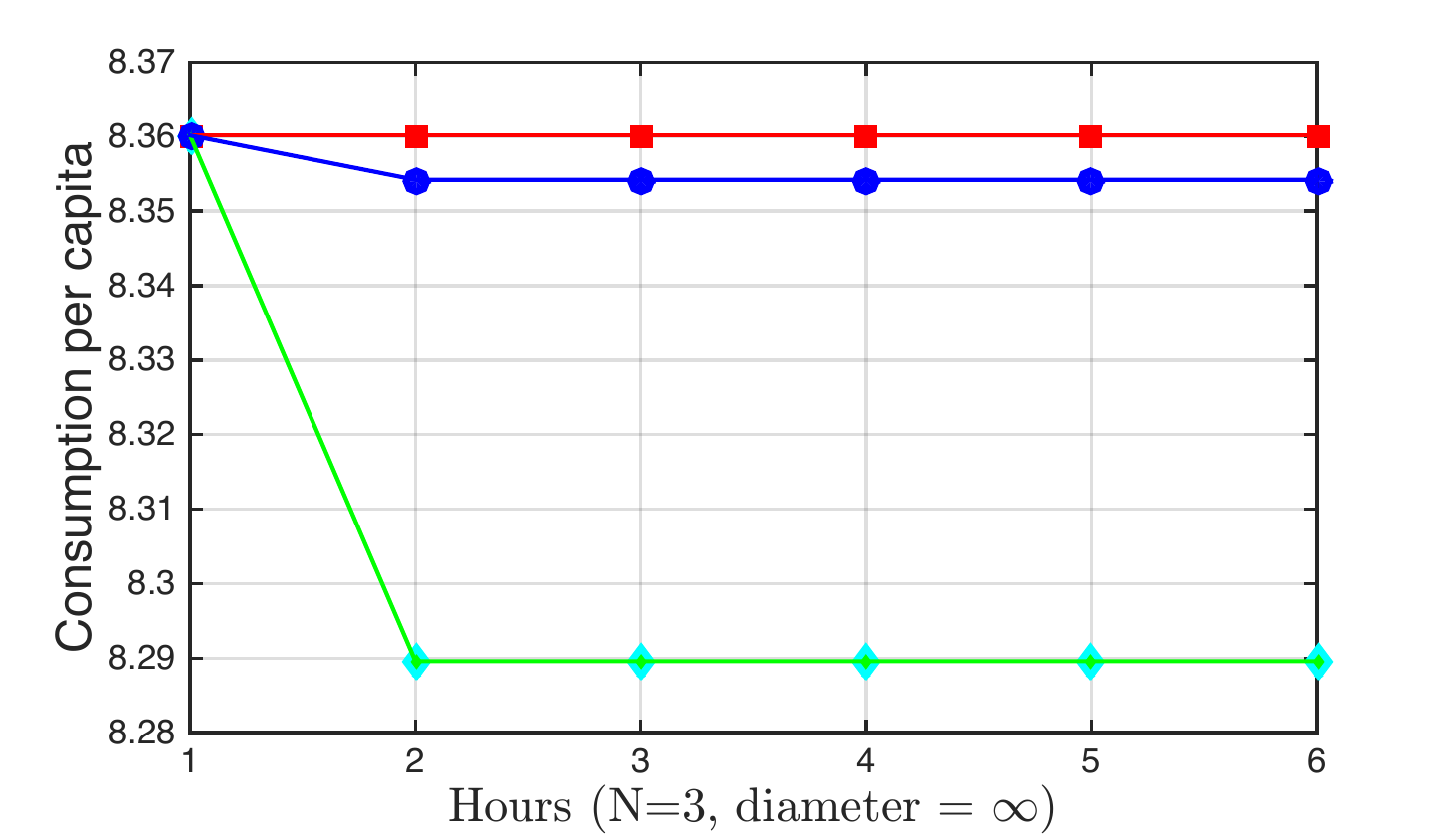}
&\includegraphics[width=0.47\linewidth]{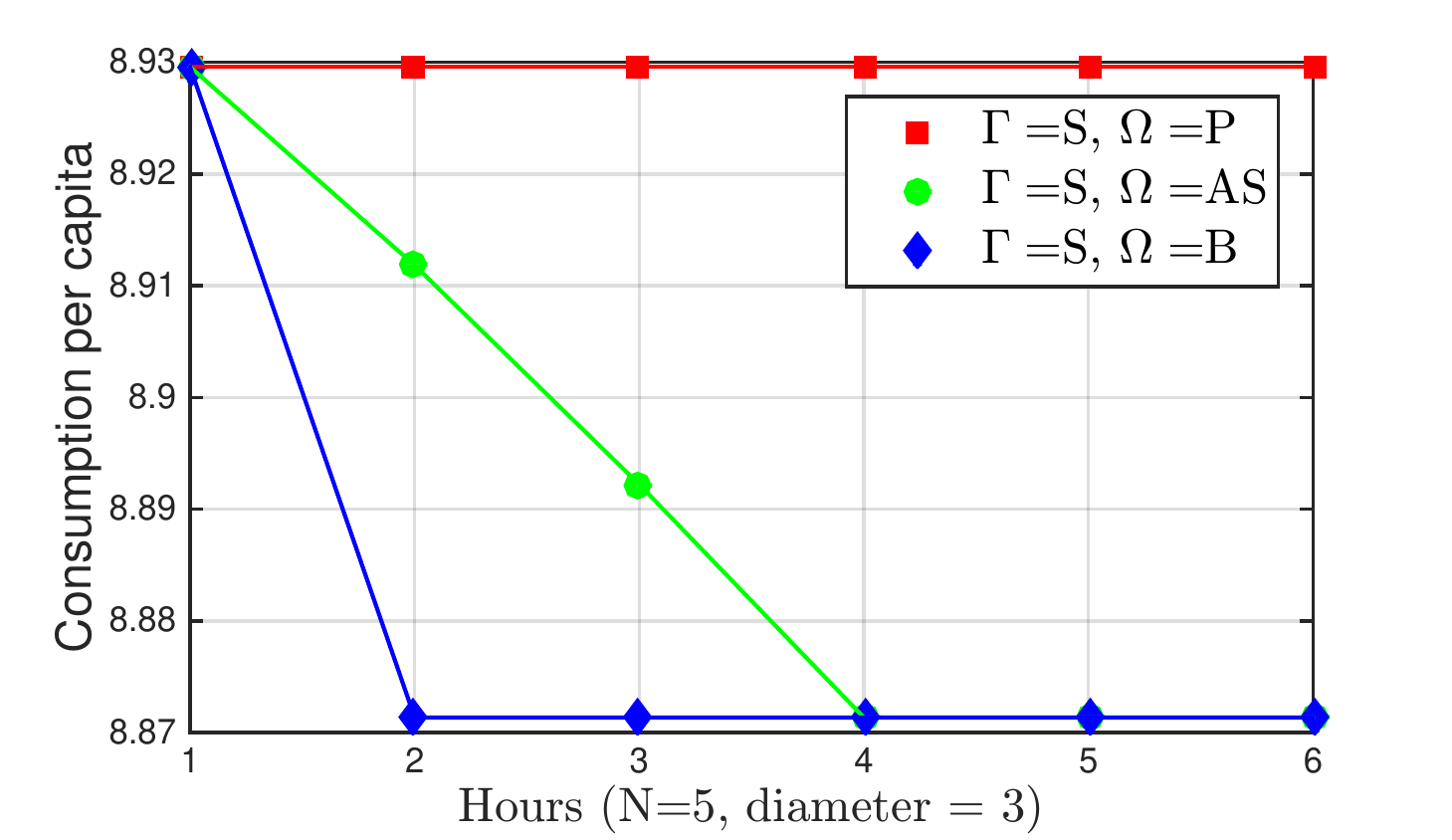}\\
    \fontsize{7}{12}\selectfont (a)
      	       & \fontsize{7}{12}\selectfont (b)\\
\includegraphics[width=0.47\linewidth]{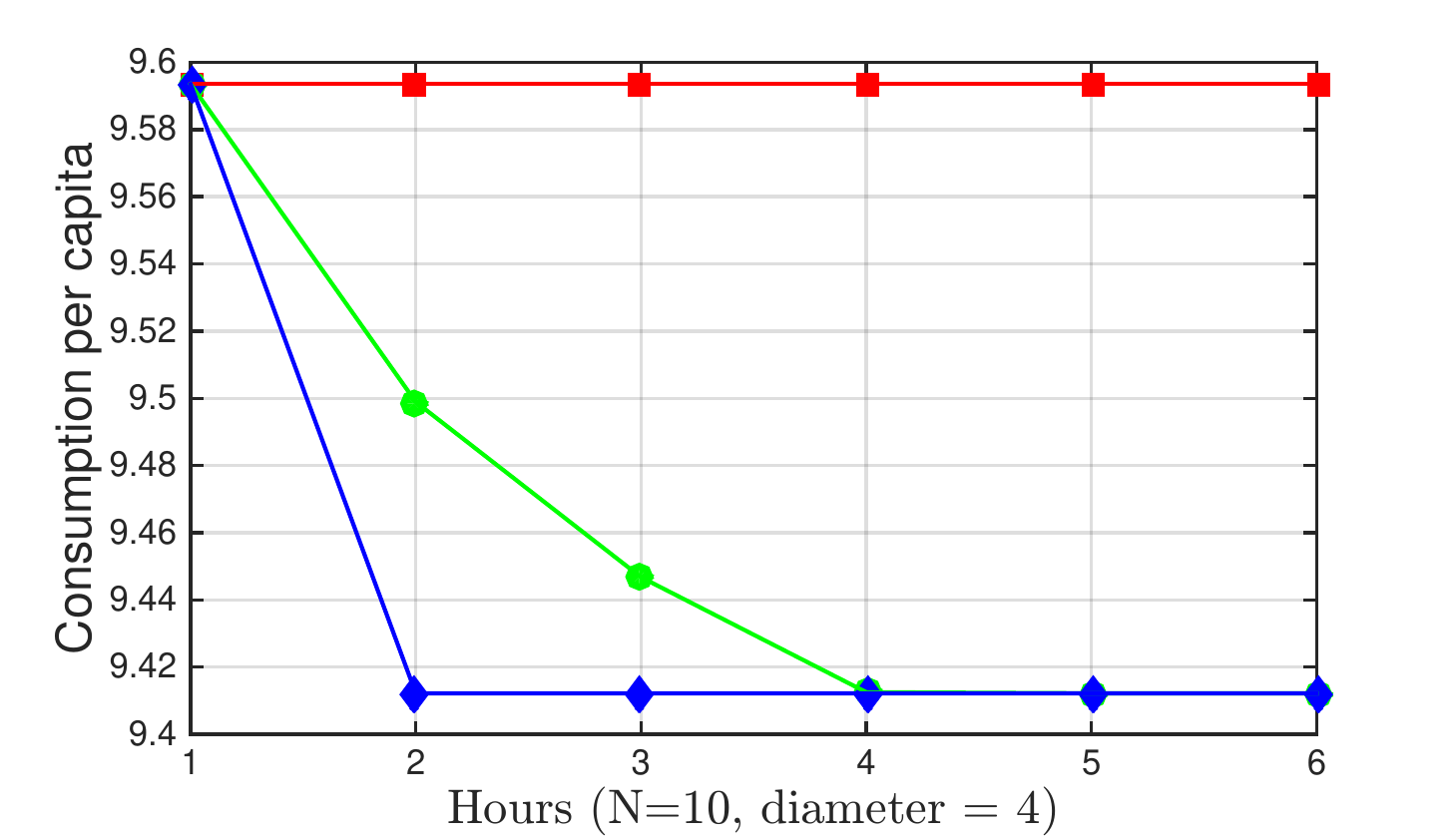} 
&\includegraphics[width=0.47\linewidth]{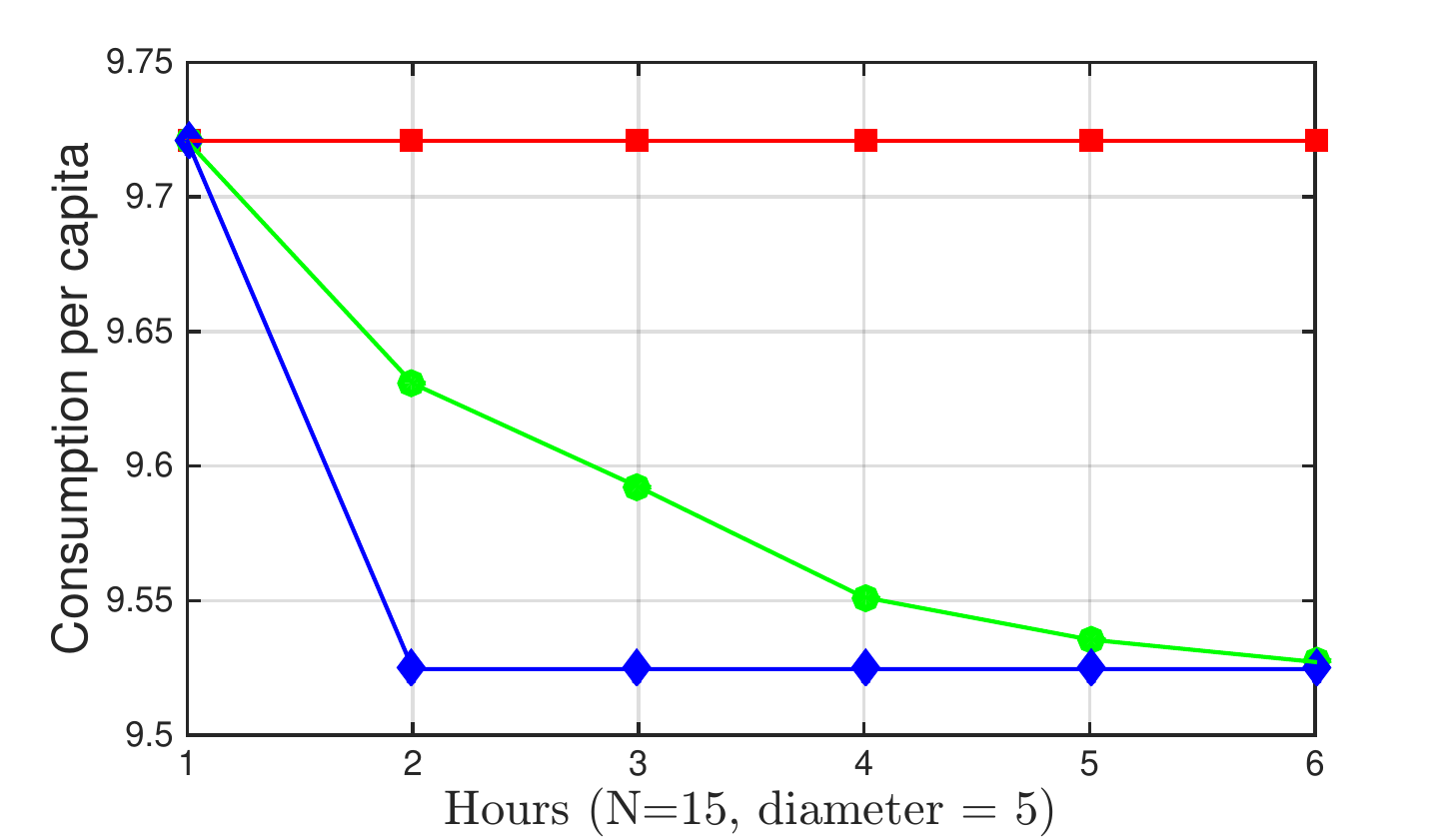} \\
      \fontsize{7}{12}\selectfont (c) 
      &    \fontsize{7}{12}\selectfont (d)\vspace{-2mm}
 \end{tabular}
\caption{Total consumption over time for $\Gamma = $S and $\Omega \in $ \{P, AS, B\} for $N =\{3,5,10,15\}$ population size. There is a single time zone $\ccalT$ which lasts for $H=5$ hours. The cost function constant $\kappa_h = 1$ for $h \in \ccalT$. The price policy parameter is chosen as $\gamma_h = 1.2 \$/\textrm{kWh}^2$ for all time slots. The communication network is determined by randomly placing $N$ individuals on a $3$-mile$\times 5$-mile area and connecting them if they are closer than the threshold connectivity of 2 miles. The diameter number under each figure indicates the network diameter. We let the decay parameter in \eqref{utility_i} be $\alpha_h = 1$ for $h \in \ccalT$. The mean of the preferences $g_i$ is equal to $30$ for $i \in \ccalN$. We let the standard deviation of the preference to be identical for all consumers as $\sigma_{ii} = 4$ and we assume preferences are uncorrelated, $\sigma_{ij}=0$. We let the renewable power term $\omega$ be normal-distributed with mean $\bar\omega_h = 0$ and variance $\sigma_{\omega} = 2$. When the network is connected, AS converges to the B in the number of steps equal to the network diameter. }
\vspace{-5mm}
\label{Performance_metrics}
\end{figure}

\begin{corollary}
Consider the complete information model with $\sigma$-correlated preferences. We have the variance of $L^\Gamma/N$ as 
\begin{equation} \label{eq_variance_demand_complete}
Var(L^\Gamma/N) =  \frac{N-1}{N^2}(a^\Gamma)^2 ,
\end{equation}
where $a^\Gamma$ is as defined in Proposition \ref{cor_BNE_characterization_complete}.
\end{corollary}

\begin{myproof}
From the definition of the variance of normalized demand, 
\begin{align}
Var&(L^\Gamma/N) := (E\big[(\sum_{j=1}^Ns_i^\Gamma(I_i^B))^2\big] - E\big[(\sum_{j=1}^Ns_i^\Gamma(I_i^B))\big]^2)/N^2 \nonumber\\
&= E\bigg[\bigg(\sum_{j=1}^N a^\Gamma (g_i - \frac{1}{N}\sum_{j=1}^N g_j)\bigg)^2\bigg]/N^2 \nonumber \\
&= \frac{(a^\Gamma)^2}{N^2} \bigg(N E\bigg[  (g_i - \frac{1}{N}\sum_{j=1}^N g_j)^2\bigg]  \nonumber \\
&+ N(N-1) E\bigg[(g_i - \frac{1}{N}\sum_{j=1}^N g_j)(g_k - \frac{1}{N}\sum_{j=1}^N g_j)\bigg]\bigg). \label{eq_variance_intermediate}
\end{align}
In the second step we substitute the consumption decisions of individuals from \eqref{eq_explicit_broadcast} and cancel out the mean of consumption squared. The third step expands the quadratic sum where we have $k \neq i$ in the last equality.
Now consider the first expectation inside the last equality above: 
\begin{equation}
E\bigg[  \bigg(g_i - \frac{1}{N}\sum_{j=1}^N g_j\bigg)^2\bigg] = \frac{N-1}{N}(1-\sigma).
\end{equation}
The second expectation inside \eqref{eq_variance_intermediate} is 
\begin{equation}
E\bigg[\bigg(g_i - \frac{1}{N}\sum_{j=1}^N g_j\bigg)\bigg(g_k - \frac{1}{N}\sum_{j=1}^N g_j\bigg)\bigg] = \frac{1}{N}\sigma.
\end{equation}
The result follows by substituting the above two identities into \eqref{eq_variance_intermediate} and then simplifying the terms.
\end{myproof}
We have the following observations for the variance of demand in the complete information case. 
\begin{corollary}
Consider the complete information model with $\sigma$-correlated preferences. For large number of consumers, $N$, 
\begin{enumerate}
\item $Var(L^\Gamma/N) \approx 0 $ for any model $\Gamma = \{{\rm S, U, W}\}$.
\item The change of variance with respect to $\sigma$ is zero.  
\end{enumerate}
\end{corollary}
\begin{myproof}
When $N$ is large, we have the constant $a^\Gamma \approx (2\alpha)^{-1}$ from Proposition \ref{cor_BNE_characterization_complete} for any $\Gamma\in {\rm \{S, U, W\}}$. The observations above follow when we substitute this approximation in \eqref{eq_variance_demand_complete}.
\end{myproof}

The results above are in sharp contrast to the observations given by items 1 and 2 of Corollary \ref{cor_variance_demand_private} for the variance of demand in the private information model. In summary, while giving information to the consumers does not affect the expected demand as per \eqref{expected_demand}, it reduces the uncertainty in demand forecasts. 

So far, we have compared the private information and complete information cases. As aforementioned, complete information is achieved after the first time step in the broadcast information exchange model during a time zone. In the action-sharing information exchange model, we expect to observe effects on demand similar to the complete information case. However, an analytical comparison between $\Omega = {\rm AS}$ and $\Omega = {\rm P}$ is not possible because we do not have a closed form solution to the consumption behavior when  $\Omega = {\rm AS}$ as per Proposition \ref{BQNG_theorem}. 

In the sequel, we numerically analyze the effects of the action-sharing information exchange model on expected consumption and its variance. Note that in the action-sharing (AS) information exchange model, we need to consider multiple time steps in a time zone to measure the effects of information. In AS, consumers follow the steps in Algorithm \ref{alg1} to compute their BNE strategies. Figs. \ref{Performance_metrics}(a)-(d) exhibit the total consumption with respect to hours for the population sizes $N = \{3,5,10,15\}$, respectively. Given a population size plot, each line corresponds to a different information exchange model for the selfish consumer behavior model -- see the legend in Fig. \ref{Performance_metrics}(b). We observe that when the network is connected (Figs. \ref{Performance_metrics}(b)-(d)), the total consumption in AS model converges to the total consumption in the B model. Furthermore, convergence time is proportional to the diameter of the network. When the network is not connected (Fig. \ref{Performance_metrics}(a)), convergence does not necessarily happen. Finally, we compare the normalized demand variance for the cases considered in Fig. \ref{Performance_metrics} in Table \ref{table:consumption}. As expected, the AS model has a demand variance that is smaller than the private information model but larger than the broadcast model. 

\begin{table}[t]
\centering
\begin{tabular}{@{}l l l l l @{}}\toprule
\multicolumn{1}{ c  }{} & \multicolumn{4}{ c }{N}\\
\cmidrule{2-5}
\multicolumn{1}{ c  }{} & 3  & 5  & 10 & 15\\ \midrule
\multicolumn{1}{ c  }{P}& 3.5   & 1.7  & 1.2 & 1.1 \\
\multicolumn{1}{ c  }{AS}& 3  & 1.1  & 0.7 & 0.6\\
\multicolumn{1}{ c  }{B}& 2.1  & 0.9  & 0.5 & 0.5\\
\bottomrule
\vspace{0pt}
\end{tabular}
\caption{Normalized demand variance with selfish consumers for the set-up in Figure \ref{Performance_metrics}}
\label{table:consumption}
\end{table}

%

\section{Effects of behavior and information exchange models on aggregate utility} \label{sec_effects_aggregate}

We begin by comparing the effects of private and complete information models on aggregate utility. In both of cases, behavior is static. Hence we focus on demand at a fixed hour and remove the sub-index $h$ from the notation. We then give an explicit characterization of the expected aggregate utility for the symmetric BNE strategies of the form given by \eqref{eq_explicit_private} or \eqref{eq_explicit_broadcast}.

\begin{lemma}\label{lemma_aggregate_utility}
For the symmetric BNE strategies of the form $s_i = a (g_i-\bar{g}) + b \bar{g}$, we have the following characterization of normalized expected aggregate utility when $\bar\omega = 0$:
\begin{equation}\label{eq_expected_utility_characterization}
\frac{E[U]}{N} = \big( b-(\gamma + \alpha) b^2 \big) \bar{g}^2 - \bigg( \frac{N-1}{N}\gamma\sigma + \frac{\gamma}{N} + \alpha\bigg)a^2 + a.
\end{equation}
\end{lemma}
\begin{myproof}
From the definition of aggregate utility $U$ in \eqref{aggregate_utility} and the definition of individual utility, we have
\begin{align} \label{eq_normalized_eu}
\frac{E[U]}{N} &= \frac{1}{N}E\bigg[\sum_{i=1}^N s_i \bigg( -\frac{\gamma}{N} \bigg( \sum_{j=1}^{N} s_j + \bar\omega\bigg)\bigg) + g_i s_i -\alpha s_i^2\bigg] \! .
\end{align}
From the above relationship there are three expectation terms we need to compute. The first term is given by
\begin{align}
E\bigg[s_i \sum_{j=1}^N s_j\bigg] &= E\bigg[ \big(a (g_i-\bar{g}) + b \bar{g}\big)  \bigg(a \sum_{j=1}^N (g_j -\bar{g}) + Nb \bar{g}\bigg)\bigg]\nonumber\\
&= a^2 \big(1 + (N-1) \sigma\big) + N b^2 \bar{g}^2.
\end{align}
The second term is 
\begin{align}
E[s_i g_i] = a + b \bar{g}^2 .
\end{align}
The third expectation is 
\begin{align}
E[s_i^2] &= E[(a (g_i-\bar{g}) + b (\bar{g}))^2]  = a^2 + b^2 \bar{g}^2.
\end{align}
Combining the three terms above in \eqref{eq_normalized_eu}, we have
\begin{align}
\frac{E[U]}{N}  &= \frac{1}{N}\bigg(-N\frac{\gamma}{N}\big(a^2 (1 + (N-1) \sigma) + N b^2 \bar{g}^2\big) \nonumber \\
&+N (a + b \bar{g}^2) - \alpha N (a^2 + b^2 \bar{g}^2)\bigg).
\end{align}
Reorganizing and simplifying gives the desired result.
\end{myproof}

We compare the effects of private and complete information exchange models on aggregate utility next. 
\begin{corollary}\label{cor_aggregate_utility_information}
Consider the private and complete information models for $\sigma$-correlated preferences for large number of consumers, $N$.
\begin{enumerate}
\item The expected normalized aggregate utility is larger in the private information model than in the complete information model. 
\item The difference in expected normalized aggregate utilities between the two information models becomes negligible as $\bar{g}$ increases. 
\end{enumerate}
\end{corollary}
\begin{myproof}
For large $N$ in private information case, $a^\Gamma \approx (N \lambda^\Gamma \sigma + 2 \alpha)^{-1}$ from Proposition \ref{cor_BNE_characterization_private}, where $\lambda^\Gamma$ is as defined in  Lemma \ref{best_response_lemma}. For large $N$ in the complete information case, $a^\Gamma \approx (2\alpha)^{-1}$ from Proposition \ref{cor_BNE_characterization_complete} for any $\Gamma$. In addition, the $b^\Gamma$ term is the same for both private and complete information models. The first observation is established by substituting these constants for the BNE behavior of both private \eqref{eq_explicit_private} and complete information  \eqref{eq_explicit_broadcast} models in \eqref{eq_expected_utility_characterization} and taking the difference. The second relationship is deduced by noting that when $\bar{g}$ increases, while the expected utility in \eqref{eq_expected_utility_characterization} increases, the difference between the two models remains the same because the constant $b$ is the same for both models.
\end{myproof}

The results above imply that at high mean consumption preference, the broadcast information has negligible effect on aggregate utility. We now compare the effects of behavior models.
\begin{corollary}
Consider the private and complete information models for $\sigma$-correlated preferences. We have the following relations in terms of the normalized expected aggregate utility for large $N$ and $\bar{g}$ for both information models:
\begin{enumerate}
\item  The ratio of expected aggregate utility when consumers act selfish versus altruistic is given by
\begin{equation}\label{eq_ratio_expected_aggregate_utility}
\frac{E[U^S]}{E[U^U]}= \frac{4 \alpha^2 + 4\alpha \gamma}{4 \alpha^2 + 4\alpha \gamma + \gamma^2}.
\end{equation}
That is, altruist behavior achieves a strictly higher aggregate utility.
\item The ratio of expected aggregate utility when consumers act as welfare maximizers versus altruistic is given by
\begin{equation}
\frac{E[U^W]}{E[U^U]}= \frac{(\gamma+\alpha)(2\kappa + \alpha - \gamma)}{(\kappa+\alpha)^2},
\end{equation}
which is strictly less than 1 for any parameter value $\gamma \neq \kappa$.
\end{enumerate}
\end{corollary}
\begin{myproof}
Note that for large $N$, we have $b^{\textrm{S}} \approx (\gamma + 2 \alpha)^{-1}$, $b^{\textrm{U}} \approx (2(\gamma + \alpha))^{-1}$ and $b^{\textrm{W}} \approx (2(\kappa + \alpha))^{-1}$ for both the private and complete information models. Proof follows by substituting the related constants of the behavior models in \eqref{eq_expected_utility_characterization} and only considering terms that multiply $\bar{g}^2$. 
\end{myproof}

As expected altruistic behavior model attains a higher aggregate utility than any other consumer behavior model regardless of the information exchange model. Furthermore, the ratio of increase in \eqref{eq_ratio_expected_aggregate_utility} when compared to selfish behavior increases as the price policy parameter $\gamma$ increases. We next consider the impact of changes in $\sigma$-correlation on the expected aggregate utility. 

\begin{corollary}
Consider the private and complete information models for $\sigma$-correlated preferences. 
\begin{enumerate}
{\item When information is private and $N$ is large, the sensitivity of expected aggregate utility to correlation constant $\sigma$ is negative for models $\Gamma = \{{\rm S, U}\}$ and for $\Gamma = {\rm W}$ when $\kappa >  \gamma/2$. In addition, the decrease in aggregate utility with respect to $\sigma$ slows as $\sigma$ grows. }
\item When information is complete and $N$ is large, the sensitivity of expected aggregate utility is given by
\begin{equation}
\frac{\partial E[U]/N}{\partial \sigma} = -\frac{\gamma}{4 \alpha^2} < 0.
\end{equation}
\end{enumerate}
\end{corollary}
\begin{myproof}
Note that $b$ terms in \eqref{eq_expected_utility_characterization} do not depend on $\sigma$. The results follow by substituting the values of $a^\Gamma$ terms when $N$ is large for the corresponding information model in \eqref{eq_expected_utility_characterization} and then taking the derivative with respect to $\sigma$. 
\end{myproof}

The above results show that the increasing correlation among user consumption preferences decreases aggregate user utility irrespective of the information model. 

\begin{figure}[!t]
\centering
\begin{tabular}{c} 
\includegraphics[width=0.9\linewidth]{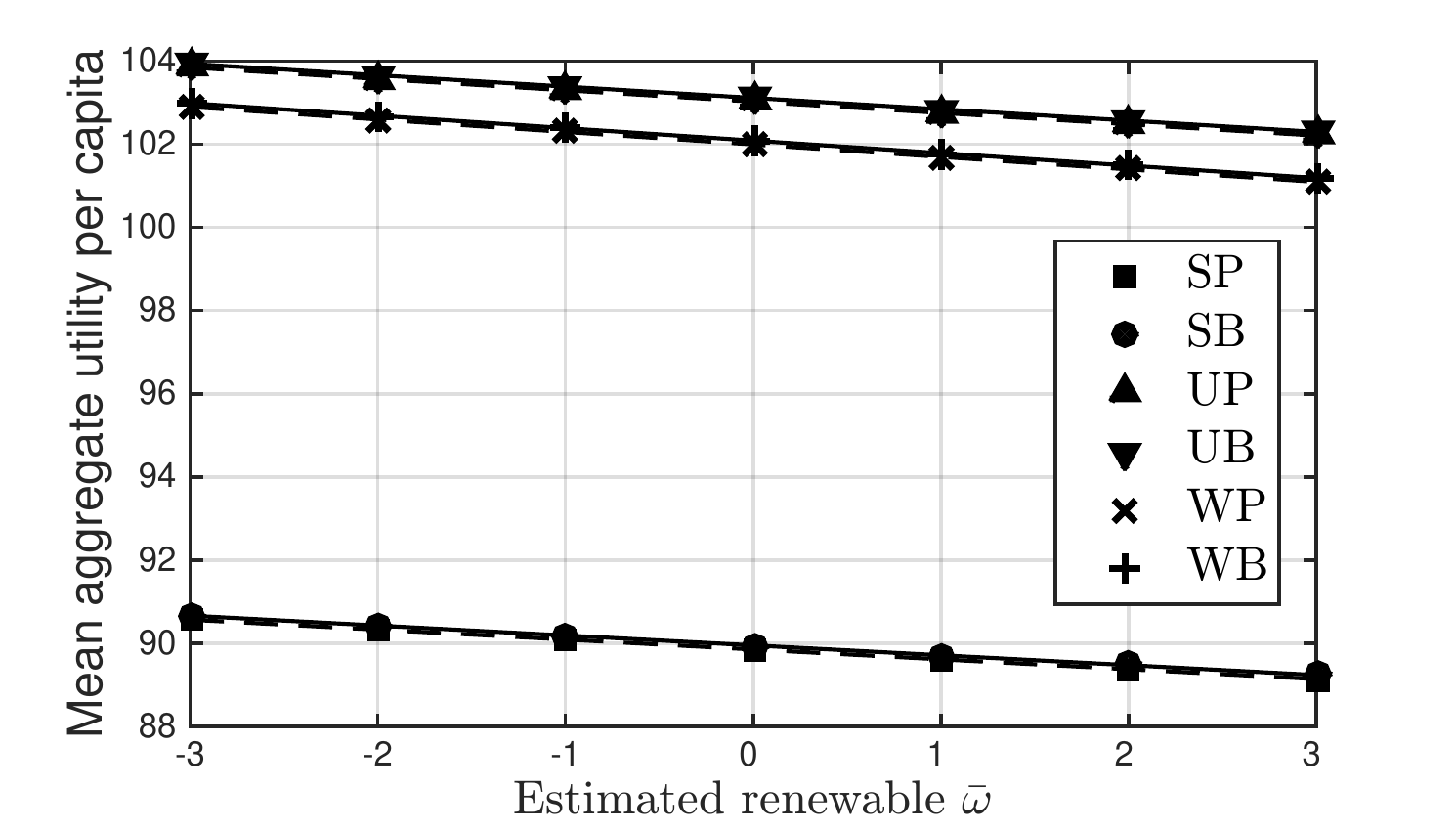}
	  \end{tabular}
\caption{Effect of mean estimate of renewable energy $\bar\omega$ on aggregate utility per capita $E{U}/N$. We let $N=30$. The renewable term $\bar\omega$ takes values in $\{-3,-2,-1,0,1,2,3\}$ and the preference correlation is fixed at $\sigma_{ij} = 2.4$. The other constants are as defined in Fig. \ref{Performance_metrics}. We consider 100 instantiations of the random variables $\bbg$ and $\omega$. We compute the expected values of aggregate utility by taking an average of all runs for a given $\bar\omega$. 
}
\label{aggregate_utility_omega}
\end{figure}

Lastly, we consider the effect of the reported mean estimate of the renewable energy term $\bar \omega$ in \eqref{price} on aggregate utility. In Fig. \ref{aggregate_utility_omega}, we plot normalized expected aggregate utility per capita $E[U]/N$ with respect to $\bar\omega \in [-3,3]$. We observe private (dashed line) and complete (solid line) information models have negligible difference in $E[U]/N$ for a given behavior model as per Corollary \ref{cor_aggregate_utility_information}. Aggregate utility decreases for all behavior and information exchange models as $\bar\omega$ increases. While the slope of decrease is the same for selfish and aggregate utility maximizers, this decrease is slightly faster for welfare maximizers. 

We compared the effects of behavior models in the context of the two static information exchange models. Results indicate that the expected aggregate utility is not affected by the information that users may be given. We also explicitly characterized the extent of the effect of consumer behavior models on aggregate utility, and concluded that consumer behavior models are the primary determinant of expected aggregate utility. 
We confirm this intuition by considering the average aggregate utility from an ensemble of runs in the set-up of Fig. \ref{Performance_metrics}. 

%

\section{Effects of behavior and information exchange models on welfare}\label{sec_effects_welfare}

We follow a similar path as the previous section. We focus on the two extreme information models, private and complete, when consumers have static information. Hence, we remove the time sub-index. The following result provides an explicit characterization of the expected welfare for symmetric BNE strategies. 

\begin{lemma} \label{lemma_welfare}
For the symmetric BNE strategies of the form $s_i = a (g_i-\bar{g}) + b \bar{g}$, we have the following characterization of normalized expected welfare when $\bar\omega = 0$: 
\begin{equation}\label{eq_expected_welfare_characterization}
\frac{E[W]}{N} = \big(b-(\kappa + \alpha) b^2 \big) \bar{g}^2 - \bigg( \frac{N-1}{N}\kappa\sigma + \frac{\kappa}{N} + \alpha\bigg)a^2 + a.
\end{equation}
\end{lemma}
\begin{myproof}
Recall the definition of welfare $W$ in \eqref{welfare} and note that $E[W]/N$ is equal to \eqref{eq_normalized_eu} when we replace $\gamma$ with $\kappa$ and let $\bar\omega = 0$ in \eqref{eq_normalized_eu}. The rest follows along the same lines as in the proof of Lemma \ref{lemma_aggregate_utility}. 
\end{myproof}

The effects of information models, private or complete, are similar to the results for aggregate utility in Corollary \ref{cor_aggregate_utility_information} as the functional form of welfare in \eqref{eq_expected_welfare_characterization} is identical. That is, for large $N$, the private information model is slightly preferable. If the mean preference $\bar{g}$ is large in comparison to the decay parameter $\alpha$, the information models have a negligible effect on welfare. Next, we compare the impact of consumer behavior models.

\begin{corollary}
Consider the private and complete information models for $\sigma$-correlated preferences. We have the following relationships in terms of the normalized expected welfare for large $N$ and $\bar{g}$ in both information models:
\begin{enumerate}
\item The ratio of expected welfare when consumers act selfish versus welfare-maximizing is 
\begin{equation} \label{eq_selfish_welfare}
\frac{E[W^S]}{E[W^W]}= \frac{4 \alpha^2 + 4\alpha \gamma+ 4 \kappa (\gamma - \kappa)}{4 \alpha^2 + 4\alpha \gamma + \gamma^2}.
\end{equation}
That is, welfare-maximizing behavior achieves a strictly higher welfare except when they are equal at $\gamma = 2 \kappa$.
\item The ratio of expected welfare when consumers act altruistic versus welfare-maximizing is 
\begin{equation}\label{eq_altruistic_welfare}
\frac{E[W^U]}{E[W^W]}= \frac{(\kappa+\alpha)(2\gamma + \alpha - \kappa)}{(\gamma+\alpha)^2},
\end{equation}
which is strictly less than 1 for any parameter value except when welfares are equal at $\gamma = \kappa $.
\end{enumerate}
\end{corollary}
\begin{myproof}
Note that for large $N$, we have $b^{\textrm{S}} \approx (\gamma + 2 \alpha)^{-1}$, $b^{\textrm{U}} \approx (2(\gamma + \alpha))^{-1}$ and $b^{\textrm{W}} \approx (2(\kappa + \alpha))^{-1}$ for both the private and complete information models. Proof follows by substituting the related constants of the behavior models in \eqref{eq_expected_welfare_characterization} and only considering terms that multiply $\bar{g}^2$. To prove that \eqref{eq_selfish_welfare} is less than 1, we need to show that $\gamma^2 > 4 \kappa (\gamma - \kappa)$. Observe that this relationship is equivalent to $(\gamma - 2 \kappa)^2>0$, which is true except for $\gamma = 2 \kappa$. Similarly, showing that \eqref{eq_altruistic_welfare} is less than 1 amounts to $2 \kappa \gamma - \gamma^2 < \kappa^2$. This is equivalent to  $(\kappa-\gamma)^2 > 0$, which is true if $\kappa \neq \gamma$. 
\end{myproof}

As expected, welfare-maximizing behavior attains a higher welfare than any other consumer behavior model regardless of the information exchange model. Furthermore, the ratio of increase in \eqref{eq_selfish_welfare}, when compared to selfish behavior, grows as the price policy parameter $\gamma$ increases. Next we consider the effect of $\sigma$-correlation on expected welfare.

\begin{corollary}
Consider the private and complete information models for $\sigma$-correlated preferences. 
\begin{enumerate}
{\item When information is private and $N$ is large, the sensitivity of expected welfare to correlation constant $\sigma$ is negative if $\alpha > \gamma$ and $\gamma > \kappa$. In addition, the decrease in welfare with respect to $\sigma$ slows as $\sigma$ grows. }
\item When information is complete and $N$ is large, the sensitivity of welfare is given by
\begin{equation}
\frac{\partial E[W]/N}{\partial \sigma} = -\frac{\kappa}{4 \alpha^2} < 0.
\end{equation}
\end{enumerate}
\end{corollary}
\begin{myproof}
Note that the $b$ terms in \eqref{eq_expected_welfare_characterization} do not depend on $\sigma$. The results follow by substituting the values of $a^\Gamma$ terms when $N$ is large for the corresponding information model in \eqref{eq_expected_welfare_characterization} and then taking the derivative with respect to $\sigma$. 
\end{myproof}

The above results show that increasing correlation among user consumption preferences decreases welfare irrespective of the information model. 

\begin{figure}[!t]
\centering
\begin{tabular}{c} 
\includegraphics[width=0.9\linewidth]{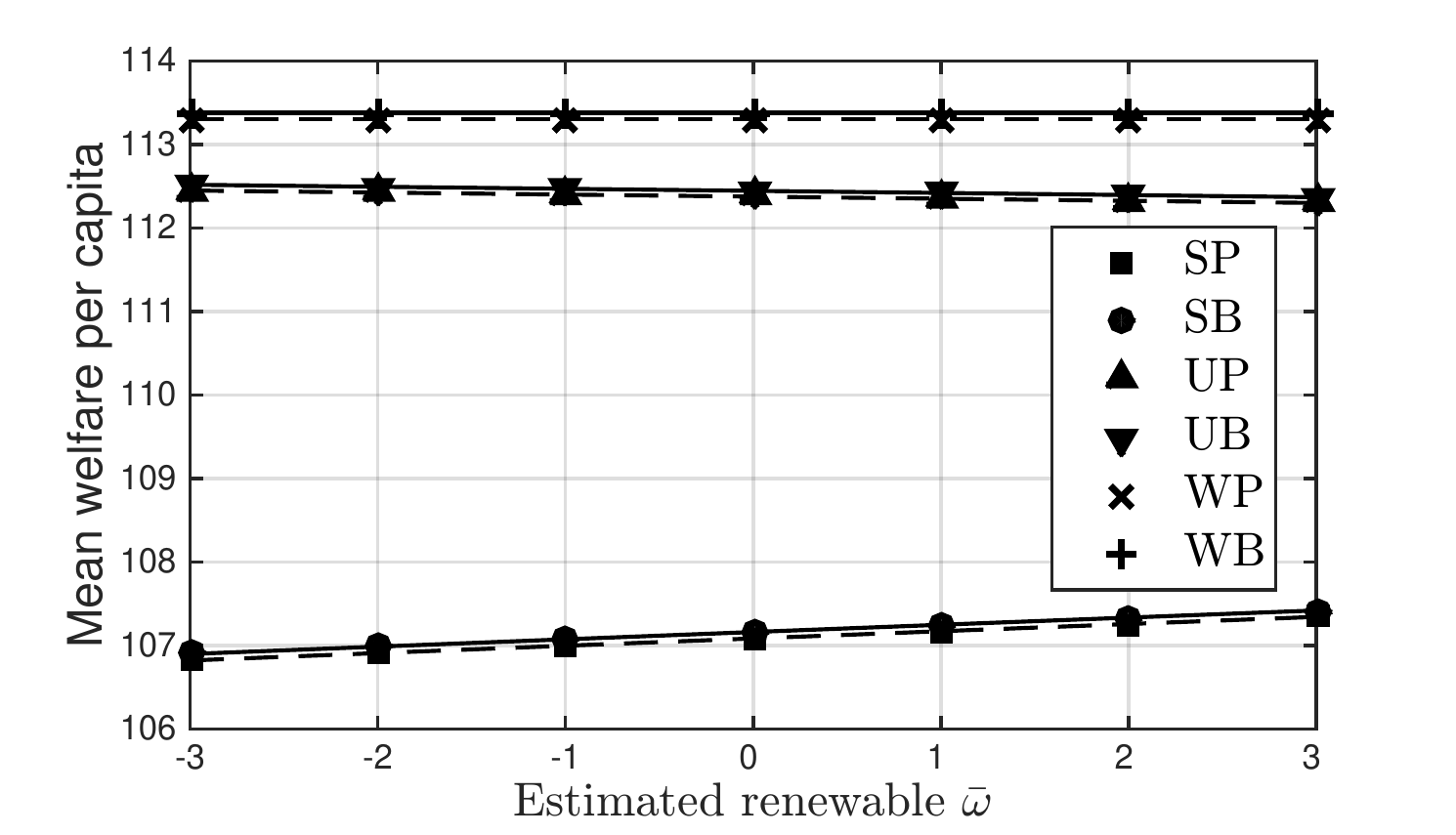}
	  \end{tabular}
\caption{Effect of mean estimate of renewable energy $\bar\omega$ on welfare per capita $E{W}/N$. The renewable term $\bar\omega$ takes values in $\{-2,-1,0,1,2\}$ and the correlation coefficient is fixed at $\sigma_{ij} = 2.4$. The other constants are as defined in Fig. \ref{Performance_metrics}. We consider 100 instantiations of the random variables $\bbg$ and $\omega$. }
\label{welfare_omega}
\end{figure}

Lastly, we consider the effect of reported mean estimate of the renewable energy term $\bar \omega$ in \eqref{price} on welfare. In Fig. \ref{welfare_omega}, we plot expected normalized welfare $E[W]/N$ with respect to $\bar\omega \in [-3,3]$.  We observe private (dashed lines) and complete (solid lines) information models have negligible differences in $E[W]/N$ for a given behavior model. $E[W]/N$ does not change for welfare maximizers because users do not respond to changes is $\bar\omega$. $E[W]/N$ increases for selfish users while it decreases for aggregate utility maximizers. To see why, recall Corollary \ref{corollary_expected_consumption} from which we have $E[s_i^{S}] > E [s_i^{W}] > E[s_i^{U}]$ for $\kappa<\gamma< 2 \kappa$. When $\bar\omega$ increases, $E[s_i^{S}]$ decreases by \eqref{expected_demand} and gets closer to $E [s_i^{W}]$ causing the increase of $E[W]/N$. When $\bar\omega$ increases, $E[s_i^{U}]$ decreases and moves away from $E [s_i^{W}]$ causing the decrease of $E[W]/N$. Considering the drop in expected aggregate utility of selfish users in Fig. \ref{aggregate_utility_omega}, increasing $\bar\omega$ implies a higher revenue for the operator when users are selfish because welfare is the sum of aggregate utility and net revenue in \eqref{welfare}. 

We compared the effects of behavior models in the context of the two static information exchange models. In sum, the results indicate that the expected welfare is not affected by the information that users may be given. We also explicitly characterized the extent to which consumer behavior models affected expected welfare. Since the latter are the primary determinants of expected welfare, we expect the action-sharing information exchange model to perform similar to the private and broadcasting information exchange models. 

%% file: main.bbl

%% file: main.bbl
\begin{thebibliography}{10}
\providecommand{\url}[1]{#1}
\csname url@samestyle\endcsname
\providecommand{\newblock}{\relax}
\providecommand{\bibinfo}[2]{#2}
\providecommand{\BIBentrySTDinterwordspacing}{\spaceskip=0pt\relax}
\providecommand{\BIBentryALTinterwordstretchfactor}{4}
\providecommand{\BIBentryALTinterwordspacing}{\spaceskip=\fontdimen2\font plus
\BIBentryALTinterwordstretchfactor\fontdimen3\font minus
  \fontdimen4\font\relax}
\providecommand{\BIBforeignlanguage}[2]{{%
\expandafter\ifx\csname l@#1\endcsname\relax
\typeout{** WARNING: IEEEtran.bst: No hyphenation pattern has been}%
\typeout{** loaded for the language `#1'. Using the pattern for}%
\typeout{** the default language instead.}%
\else
\language=\csname l@#1\endcsname
\fi
#2}}
\providecommand{\BIBdecl}{\relax}
\BIBdecl

\bibitem{Autonomous_DSM_SG}
A.~H. Mohsenian-Rad, V.~W. Wong, J.~Jatskevich, R.~Schober, and A.~Leon-Garcia,
  ``Autonomous demand-side management based on game-theoretic energy
  consumption scheduling for the future smart grid,'' \emph{IEEE Trans. Smart
  Grid}, vol.~1, no.~3, pp. 320--331, Dec. 2010.

\bibitem{SmartGrid_mechanism_design}
P.~Samadi, A.~H. Mohsenian-Rad, R.~Schober, and V.~W. Wong, ``Advanced demand
  side management for the future smart grid using mechanism design,''
  \emph{IEEE Trans. Smart Grid}, vol.~3, no.~3, pp. 1170--1180, Sept. 2012.

\bibitem{Jarmo_et_al}
J.~Lun\'{e}n, S.~Werner, and V.~Koivunen, ``Distributed demand-side
  optimization with load uncertainty,'' in \emph{{Int. Conf. Acoustics, Speech
  and Signal Process.}}, Vancouver, Canada, May 2012, pp. 5229--5232.

\bibitem{Xu_Schaar_2015}
J.~Xu and M.~van~der Schaar, ``{Incentive-compatible demand-side management for
  smart grids based on review strategies},'' \emph{EURASIP Journal on Advances
  in Signal Processing}, vol.~51, pp. 1--17, Dec. 2015.

\bibitem{Li_et_al_2011}
N.~Li, L.~Chen, and S.~H. Low, ``Optimal demand response based on utility
  maximization in power networks,'' in \emph{{IEEE Power and Energy Society
  General Meeting}}, July 2011, pp. 1--8.

\bibitem{Atzeni_et_al}
I.~Atzeni, L.~Ord��ez, G.~Scutari, D.~Palomar, and J.~Fonollosa,
  ``Demand-side management via distributed energy generation and storage
  optimization,'' \emph{IEEE Trans. Smart Grid}, vol.~4, no.~2, pp. 866--876,
  June 2013.

\bibitem{Yang_et_al}
P.~Yang, G.~Tang, and A.~Nehorai, ``A game-theoretic approach for optimal
  time-of-use electricity pricing,'' \emph{IEEE Trans. Power Systems}, vol.~28,
  no.~2, pp. 884--892, May 2013.

\bibitem{Depuru_et_al_2011}
S.~Depuru, L.~Wang, and V.~Devabhaktuni, ``Smart meters for power grid:
  Challenges, issues, advantages and status,'' \emph{Renewable and Sustainable
  Energy Reviews}, vol.~15, no.~6, pp. 2736--2742, Aug. 2011.

\bibitem{Wang_et_al_2014}
Y.~Wang, W.~Saad, N.~B. Mandayam, and H.~V. Poor, ``Integrating energy storage
  into the smart grid: A prospect theoretic approach,'' in \emph{IEEE Int.
  Conf. Acoustics, Speech and Signal Process.}, May 2014, pp. 7779--7783.

\bibitem{Eksin_et_al_15}
C.~Eksin, H.~Deli\c{c}, and A.~Ribeiro, ``Demand response management in smart
  grids with heterogeneous consumer preferences,'' \emph{IEEE Trans. Smart
  Grid}, vol.~6, no.~6, pp. 3082--3094, Nov. 2015.

\bibitem{Roozbehani}
M.~Roozbehani, A.~Faghih, M.~I. Ohannessian, and M.~A. Dahleh, ``The
  intertemporal utility of demand and price elasticity of consumption in power
  grids with shiftable loads,'' in \emph{{50th IEEE Conf. Dec. and Control, and
  European Control Conf.}}, Dec. 2011, pp. 1539--1544.

\bibitem{EksinEtal15c_a}
C.~Eksin, H.~Deli\c{c}, and A.~Ribeiro, ``Rational consumer behavior models in
  smart pricing,'' in \emph{{Proc. IEEE Int. Conf. Acoustics, Speech and Signal
  Process.}}, Brisbane, Australia, Apr. 2015, pp. 3167--3171.

\bibitem{Mohsenian_et_al_2011}
C.~Wu, H.~Mohsenian-Rad, J.~Huang, and A.~Y. Wang, ``Demand side management for
  wind power integration in microgrid using dynamic potential game theory,'' in
  \emph{{IEEE GLOBECOM Workshops}}, Dec. 2011, pp. 1199--1204.

\bibitem{Gan_et_al}
L.~Gan, A.~Wierman, U.~Topcu, N.~Chen, and S.~H. Low, ``Real-time deferrable
  load control: handling the uncertainties of renewable generation,'' in
  \emph{{4th Int. Conf. Future Energy Systems}}, Jan. 2013, pp. 113--124.

\bibitem{Sioshansi_Short_2009}
R.~Sioshansi and W.~Short, ``Evaluating the impacts of real time pricing on the
  usage of wind power generation,'' \emph{IEEE Trans. Power Systems}, vol.~24,
  no.~2, pp. 516--524, May 2009.

\bibitem{Papavasiliou_Oren_2014}
A.~Papavasiliou and S.~Oren, ``Large-scale integration of deferrable demand and
  renewable energy sources,'' \emph{IEEE Trans. Power Systems}, vol.~29, no.~1,
  pp. 489--499, Jan. 2014.

\bibitem{Samadi_et_al_2010}
P.~Samadi, A.-H. Mohsenian-Rad, R.~Schober, V.~W. Wong, and J.~Jatskevich,
  ``Optimal real-time pricing algorithm based on utility maximization for smart
  grid,'' in \emph{1st IEEE Int. Conf. Smart Grid Comm.}, Oct. 2010, pp.
  415--420.

\bibitem{Chakraborty_Khargonekar_2014}
P.~Chakraborty and P.~P. Khargonekar, ``A demand response game and its robust
  price of anarchy,'' in \emph{IEEE Int. Conf. Smart Grid Comm.}, Nov. 2014,
  pp. 644--649.

\bibitem{Jiang_Low_2011_b}
L.~Jiang and S.~H. Low, ``Multi-period optimal energy procurement and demand
  response in smart grid with uncertain supply,'' in \emph{{50th IEEE Conf.
  Dec. and Control, and European Control Conf.}}, Dec. 2011, pp. 4348--4353.

\bibitem{Power_operation_control}
A.~J. Wood and B.~F. Wollenberg, \emph{Power {g}eneration, {o}peration, and
  {c}ontrol}.\hskip 1em plus 0.5em minus 0.4em\relax New York, NY: John Wiley
  \& Sons, 2012.

\bibitem{Fudenberg_Tirole_1991}
D.~Fudenberg and J.~Tirole, \emph{Game {T}heory}.\hskip 1em plus 0.5em minus
  0.4em\relax Cambridge, Massachusetts 393: MIT Press, 1991.

\bibitem{Eksin_et_al_2013}
C.~Eksin, P.~Molavi, A.~Ribeiro, and A.~Jadbabaie, ``Bayesian quadratic network
  game filters,'' \emph{IEEE Trans. Signal Process.}, vol.~62, no.~9, pp.
  2250--2264, May 2014.

\bibitem{Ho_Chu}
Y.~C. Ho and K.~Chu, ``Team decision theory and information structures in
  optimal control problems: Part {I},'' \emph{IEEE Trans. on Autom. Control},
  vol.~17, no.~1, pp. 15--22, 1972.

\bibitem{Kay}
S.~Kay, \emph{Fundamentals of Statistical Signal Processing: Estimation
  Theory}, 1st~ed.\hskip 1em plus 0.5em minus 0.4em\relax Prentice Hall,
  Englewood Cliffs, New Jersey, 1993.

\bibitem{Vives_2008}
X.~Vives, \emph{Information and Learning in Markets}.\hskip 1em plus 0.5em
  minus 0.4em\relax Princeton University Press, 2008.

\bibitem{Vives_2011}
------, ``Strategic supply function competition with private information,''
  \emph{Econometrica}, vol.~79, no.~6, pp. 1919--1966, 2011.

\bibitem{Huang_et_al_2015}
Q.~Huang, M.~Roozbehani, and M.~A. Dahleh, ``Efficiency-risk tradeoffs in
  electricity markets with dynamic demand response,'' \emph{IEEE Trans. on
  Smart Grid}, vol.~6, no.~1, pp. 279--290, Jan. 2015.

\bibitem{Forouzandehmehr_et_al_2013}
N.~Forouzandehmehr, S.~M. Perlaza, Z.~Han, and H.~V. Poor, ``A satisfaction
  game for heating, ventilation and air conditioning control of smart
  buildings,'' in \emph{IEEE Global Communications Conf.}, Dec. 2013, pp.
  3164--3169.

\end{thebibliography}
